\documentclass[a4paper, 10pt]{article}

\usepackage{amsmath}
\usepackage{amsthm}
\usepackage{amssymb}
\usepackage{mathtools}

\usepackage[table, dvipsnames]{xcolor}
\usepackage[shortlabels]{enumitem}
\usepackage{booktabs}
\usepackage{chemformula}
\usepackage[ruled]{algorithm2e}
\usepackage[colorlinks=true, linkcolor=black, citecolor=ForestGreen, urlcolor=cyan]{hyperref}
\usepackage{caption}
\usepackage{subcaption}
\usepackage[capitalise, noabbrev]{cleveref}
\usepackage{authblk}
\usepackage{doi}

\crefformat{footnote}{#2\footnotemark[#1]#3}

\usepackage[symbol]{footmisc}

\usepackage[UKenglish]{babel}
\usepackage{csquotes}
\usepackage[shortcuts]{extdash}
\usepackage[useregional]{datetime2}

\usepackage{graphicx}

\usepackage{parskip}
\usepackage{microtype}
\usepackage[left=1.2cm, right=1.2cm]{geometry}
\usepackage{float} 

\usepackage[notextcomp]{stix2}
\usepackage[font=small]{caption}
\usepackage[numbers,super,comma,sort&compress]{natbib}

\title{Resilience metrics to guide back-up investments in the power system during extreme weather}

\author[1]{Aleksander Grochowicz\thanks{Corresponding author, \url{algro@dtu.dk}}}
\author[2]{Hannah C. Bloomfield}
\author[1]{Marta Victoria}

\affil[1]{Department of Wind and Energy Systems, Technical University of Denmark, 2800 Kgs. Lyngby, Denmark}
\affil[2]{Department of Civil and Geospatial Engineering, School of Engineering, Newcastle University, Newcastle upon Tyne, United Kingdom, NE1 7RU}

\date{\today}

\begin{document}


\maketitle


\subsection*{Abstract}
Security of supply is a common and important concern when integrating renewables in net-zero power systems.
Extreme weather affects both demand and supply leading to power system stress; in Europe this stress spreads continentally beyond the meteorological root cause.
We use an approach based on shadow prices to identify periods of elevated stress called system-defining events and analyse their impact on the power system.
By classifying different types of system-defining events, we identify challenges to power system operation and planning.
Crucially, we find the need for sufficient resilience back-up (power) capacities whose financial viability is precarious due to weather variability and weather-induced risk. 
Furthermore, we disentangle short- and long-term resilience challenges (from multi-day to annual scale) with distinct metrics and stress tests to incorporate both into future energy modelling assessments.
Our methodology and implementation in an open energy system model (PyPSA-Eur) can be re-applied to other systems and help researchers and policymakers in building more resilient and adequate energy systems.


\subsection*{Introduction}
Weather uncertainty poses challenges in both planning and operating energy systems. 
While short periods of low renewable generation can generate difficulties in hourly-to-daily system balancing, variability on longer time scales (e.g. years to decades) complicates optimal sizing of future energy infrastructure, potentially imperilling security of supply.
Over Europe, there is an increasing understanding of characteristics of weather risks in guaranteeing system adequacy due to rapid recent developments in open source inputs \cite{hofmann_hampp_etal_2021,pfenninger_staffell_2016,hersbachh._bellb._etal_2018, bloomfield_brayshaw_2021, copernicusclimatechangeservice_2024} and modelling workflows \cite{horsch_hofmann_etal_2018, pfenninger_pickering_2018, goke_2021,price_zeyringer_2022}. 
With the aid of high performance computing (HPC) clusters, larger datasets of weather-dependent supply and demand are passed through power system models for optimal planning under inter-annual climate variability.
Recent modelling studies include high-resolution optimisations based on multiple weather years  \cite{zeyringer_price_etal_2018,dowling_rinaldi_etal_2020,ruhnau_qvist_2022,grochowicz_greevenbroek_etal_2023,gotske_andresen_etal_2024,grochowicz_greevenbroek_etal_2024,ruggles_virguez_etal_2024,vangreevenbroek_grochowicz_etal_2025,killenberger_zielonka_etal_2025} for more robust system layouts.

System stress in highly renewable power systems tends to occur in the European winter when seasonally low solar generation coincides with cold temperatures (therefore, increased demand) but critically, also low wind speeds, leading to ``energy droughts'' \cite{vandermost_vanderwiel_etal_2024,wiel_selten_etal_2020, vanderwiel_bloomfield_etal_2019}. 
The interplay with hydropower \cite{gotske_victoria_2021,vandermost_vanderwiel_etal_2024} is also important in systems where this is prevalent.
While meteorological conditions for system stress have been extensively studied, the effects on the power system dynamics and insights on planning are not well understood \cite{craig_wohland_etal_2022,grochowicz_greevenbroek_etal_2024,kittel_roth_etal_2026}. 
System stress has predominantly been studied by either considering thresholds on availability or net load time series (``demand -- renewable generation'') \cite{bloomfield_suitters_etal_2020, kittel_schill_2026, kelder_heinrich_etal_2025, wilczak_kirk-davidoff_etal_2025} rather than through power system model simulations.
Similarly, national studies investigate the breadth of meteorological conditions that could lead to system stress in an individual country \cite{mockert_grams_etal_2023} and the potential national planning considerations \cite{killenberger_zielonka_etal_2025, bloomfield_2025}. 
However, these studies disregard the interconnected nature of the European grid, either considering countries in isolation or a fully copperplate European system.
The appraisal and shortcomings of national perspectives and pure energy meteorology have been discussed in Ref. \cite{grochowicz_greevenbroek_etal_2024} together with the introduction of an approach based on shadow prices that synthesises weather information with the power system dynamics resulting in a more accurate identification of energy droughts through \emph{system-defining events (SDEs)}.
While Grochowicz et al. \cite{grochowicz_greevenbroek_etal_2024} compiled weather patterns of energy system stress events, their inadequate representation of long-duration storage could not explore the subtleties between energy and power deficits, i.e. storage versus discharge capacity limitations.
This has large consequences on sizing of back-up capacities, which are central to contingency plans ensuring the reliability of the power system, and flexible back-ups remain a concern for viability\cite{ERAA2024}.
Thus, the impact of extreme weather on system planning and which system components contribute to increased resilience and security remained unanswered.

In this paper, we characterise four types of short-term extreme events and their impact on operation and planning of power systems, and
\begin{enumerate}
	\item show that the back-up capacities for resilience entail financial risk tied to weather variability;
	\item separate the issue of long-term resilience (sizing of sufficient generation capacities) and short-term stress tests (sizing of sufficient back-up capacities) in power system planning through adequate metrics and approaches for incorporation.
\end{enumerate}

We conduct this study with the open model PyPSA-Eur which co-optimises capacity expansion and dispatch for a net-zero European power system with high temporal (hourly) and spatial resolution (90 nodes). 
For this, we use 80 years of ERA5 weather data (1941--2021) to capture both variability and extreme events driving supply of renewable power and (temperature-dependent) electricity demand, and identify system-defining events \cite{grochowicz_greevenbroek_etal_2024} based on shadow prices.
Though our approach is computationally expensive (60 GB RAM and solving time of $80 \cdot 3$ hours in HPC), spanning such long time periods can assess risks and shed light on the worst-case conditions we can expect in net-zero power systems. 
Our methodology identifies periods and weather years that should be prioritised by energy modellers, transmission system operators, and policymakers when implementing resource adequacy assessments for Europe.
At the same time, our set-up can be adapted to other regions or contexts and lead to different valuable insights about system-defining events for the system in question.

\subsection*{Recurring energy droughts and need for resilience reserves}
\begin{figure}[H]
	\centering
	\includegraphics[width=\linewidth]{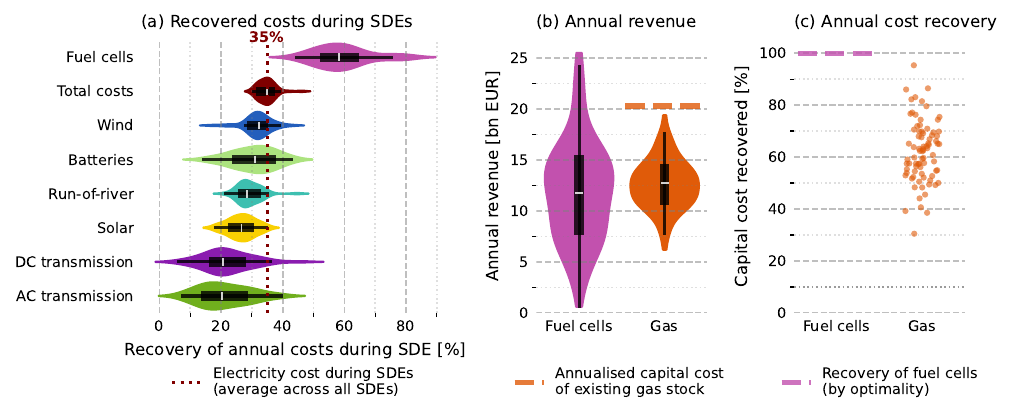}
	\caption{\textbf{Income and cost recovery of resilience back-ups are highly variable and risky.} (a) shows the extent to which different technologies recover their capital cost during SDEs. The cost recovery of fuel cells and hydrogen storage is implicitly included in the values for fuel cells. Marked with a dotted red line is the average cost inflicted on the total system (approximately a third of annual electricity cost fall into SDEs). (b) shows the annual revenue of resilience back-ups across 80 weather years: in the default scenario, fuel cells are the resilience back-ups. The values for gas come from the sensitivity analysis (99\% emission reduction) in which the existing gas stock (its capital cost marked with a dashed orange line) can be used to provide flexibility and fully replaces fuel cells entirely. (c) shows how much of annualised capital cost can be recovered over the 80 weather years. As fuel cell capacities are endogenously optimised, they always recover their costs. Gas power plants (in the sensitivity scenarios) are assumed to be kept at current capacities, and never recover their costs. In (a) and (b), the whiskers show the minimum and maximum values, whereas the boxes show the median (in white) and the lower and upper quartiles. The statistics in (a) are based on 41 system-defining events (and their annual costs), whereas the statistics in (b) and the stripplot in (c) are based on 80 weather years.}
	\label{fig:costs}
\end{figure}

\begin{figure}[H]
      \centering
      \vspace*{-2.5cm}
      \begin{subfigure}{\linewidth}
          \centering
          \includegraphics[width=\linewidth]{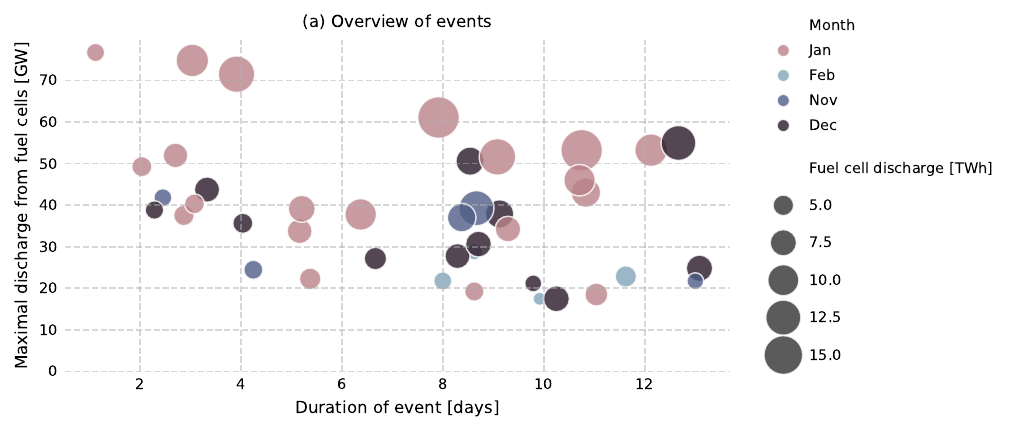}
      \end{subfigure}
      \begin{subfigure}{\linewidth}
          \centering
          \includegraphics[width=\linewidth]{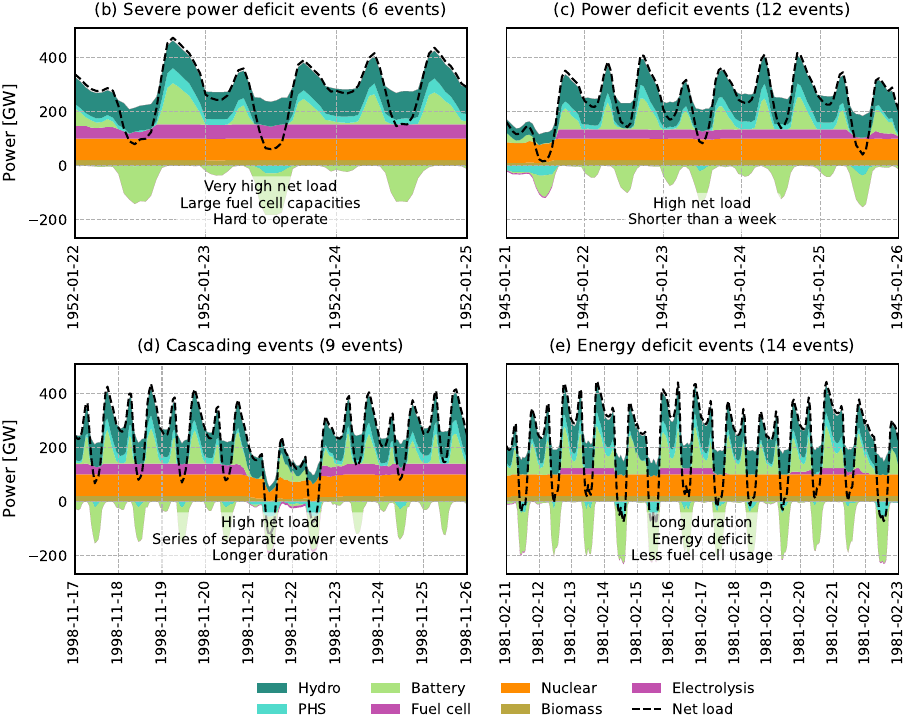}
      \end{subfigure}

      \caption{\textbf{System-defining events vary in duration and severity.}
      (a) Overview of identified SDEs according to intensity (assessed with maximal fuel cell discharge), duration (in days), and occurrence during the year (by month). The bubble size shows total hydrogen
  storage discharge during the event, indicating the need for energy storage capacity and relating power and energy deficit. (b)-(e) Four clusters of SDEs with an example event each showing dispatch stacks. See \cref{tab:clusters} and \cref{fig:kpis-clusters} for more detailed data on the identified clusters. }
      \label{fig:clusters}
  \end{figure}

Planning for increasing wind and solar penetration aims to ensure reliable operations and adequacy \cite{schmitz_flachsbarth_etal_2025b} for decades to come. 
We follow Grochowicz et al.\cite{grochowicz_greevenbroek_etal_2024} in identifying power system stress through SDEs: periods during which a significant portion of the energy infrastructure build-out is determined, measured by a rapidly accumulated cost (Methods) \cite{furmann_gotske_etal_2025, demarco_mannhardt_etal_2025}.
Although this cost includes expansion of renewable generation as well as transmission and energy storage, it is most striking for resilience provision during renewable energy droughts (\cref{fig:costs}(a)).
This differs from the straight-forward interpretation of shadow prices in pure dispatch optimisation models \cite{su_kern_etal_2020, brown_neumann_etal_2025} as electricity prices, yet results in similar timing of price spikes.
In Europe, the most challenging renewable energy droughts (SDEs account for $\sim$1\% of all hours) are restricted between November and February (\cref{fig:clusters}(a)) when solar generation is at its lowest. 
SDEs occur on average every other year, so many events are expected to fall into the lifetime of common energy infrastructure. 

During SDEs, adverse weather conditions (high pressure systems leading to low wind speeds and low temperatures \cite{grochowicz_greevenbroek_etal_2024}) reduce the seasonal share of wind power from about 50\% to 15--30\% of generation in Europe, whilst lower temperatures elevate electricity demand (Supplementary Figure S1).
Though system-defining events are more quantitatively defined than the common, more loose usage of ``energy droughts'' or ``dunkelflauten'', for all of these, it is true that the deficit between load and renewable generation may, but need not, stem from a compounding effect on both the supply and demand sides.
This deficit is detected through ``net load'' (demand -- net renewables) spikes during extreme periods and as high as 540~GW in our study (with load peaking below 700~GW; \cref{fig:combined_gen_analysis}). 
It serves as a precise measure of how much flexible generation is needed in order to compensate the shortfall of renewable generation during extreme events \cite{biewald_cozian_etal_2025,grochowicz_greevenbroek_etal_2024}.

\begin{figure}[H]
    \centering
    \includegraphics[width=\linewidth,]{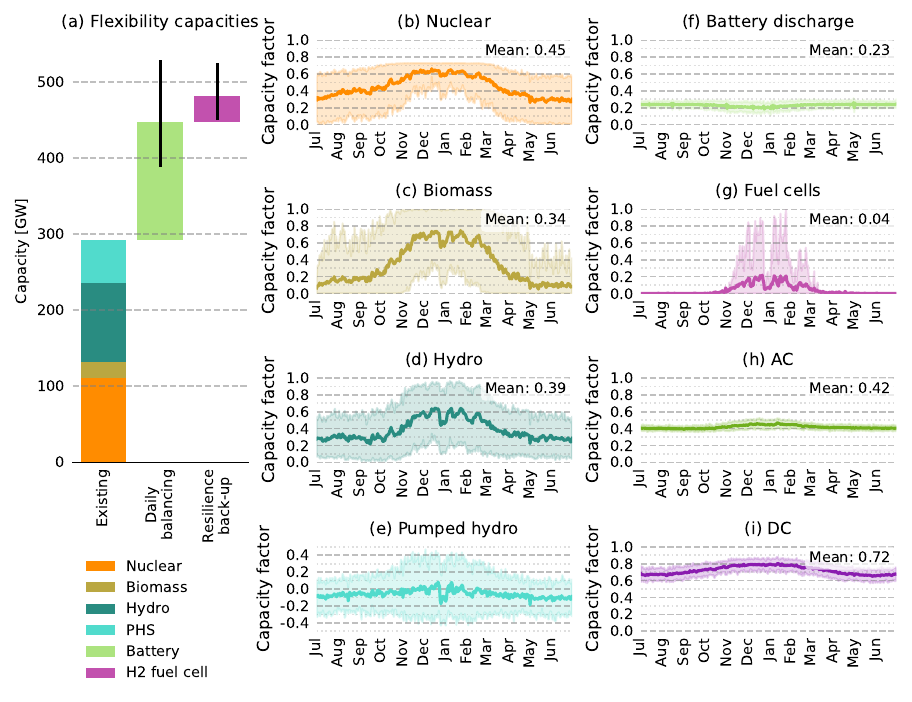}
    \caption{\textbf{Three types of flexibility (existing dispatchable capacities, batteries, resilience back-ups) stabilise the system during the winter.} Installed capacities and annual generation profiles of flexibility providers. (a) shows the capacities of the three categories of flexible generation (existing dispatch, daily balancing, resilience back-up) with uncertainty bars showing the spread of capacities (minimal and maximal capacities) across different weather years for batteries and fuel cells, with the average capacity across 80 weather years indicated by the stacked bar. Capacities for nuclear, hydropower, pumped hydro storage (PHS) and biomass power generation are not optimised, but assumed exogenously according to their values in 2025 (Methods). (b) - (i) show the mean annual capacity factor profiles for nuclear, biomass, hydropower, PHS, battery dischargers, fuel cells, as well as AC and DC transmission. The 10- and 90-percentiles are marked with the area shaded in between.}
    \label{fig:capacities}
\end{figure}

Through our set-up, we classify three types of flexibility providers apart from transmission (\cref{fig:capacities}): 
first, \emph{existing dispatchable} capacities (nuclear, biomass, hydropower and pumped hydro storage) of 300~GW are included in the system for historical reasons, but remain insufficient for approximately 1000h/year (Supplementary Figure S2).
\cref{fig:capacities}(b)-(e) show an uptick in winter utilisation of all four technologies, particularly during SDEs when the full stack of hydropower compensates renewable generation deficits partly. 
While the compensation during SDEs (or generally more difficult periods) is hinted at by the percentile shadings in \cref{fig:capacities}, the operation of hydropower and the other technologies during SDEs is best seen in \cref{fig:combined_gen_analysis}.
Second, battery capacities provide \emph{daily balancing} of 100--230 GW (depending on the weather year, \cref{fig:capacities}(a)), are dimensioned in unison with low-cost solar generation.
While helpful during peak hours of SDEs (flexible discharge up to 190~GW), they cannot cover the full duration of SDEs (\cref{fig:capacities}(f)).
Generally, existing dispatchable capacities and daily balancing are insufficient for the 100--150 most challenging hours ($\sim$1\% per year), which are part of SDEs (Supplementary Figure S2 and \cref{fig:fc_duration}).
Third, \emph{resilience back-up} capacities --- here fuel cells, existing gas power plants in our sensitivity analysis --- cover extreme periods  with dispatch at full capacity less than 50--100 hours/year (\cref{fig:fc_duration}). 
We simplify our analysis by focusing on one carbon-free and one carbon-emitting technology due to the computational burden of tracking the carbon cycle or combined heat- and power (CHP) modelling through sector coupling\cite{gotske_andresen_etal_2024}.
In reality, biomass or natural gas power plants or CHP with carbon capture and storage (CCS), as well as hydrogen turbines, could also serve as resilience back-ups (addressed in the Methods).
Resilience back-ups regardless of technology are characterised by very low capacity factors (4\% annually, \cref{fig:capacities}(g)) running only during difficult periods in the winter. 
The size of these back-ups is determined by system-defining events, and fluctuates between virtually negligible capacities in 2019/20 and 77~GW in 1965/66 (Supplementary Figure S3(a)).
Thus, while the solar-wind capacity mix (providing more than 80~GW during the worst hours of SDEs; Supplementary Figure S1 and \cref{fig:combined_gen_analysis}) determines the need for daily balancing, the remaining capacities required for resilience vary, and are determined by uncertain occurrence and variable severity of SDEs in a given year.
We find four main clusters of SDEs which we characterise in a later section.
With changing technology mixes across weather, region, and system, different kinds of threats may loom larger in some designs \cite{ruggles_virguez_etal_2024}.

\subsection*{Financial risk of resilience back-ups} 

Planning resilience back-ups is challenging because optimal investment strategies across weather years differ significantly: fuel cells are already at the edge of profitability and small perturbations in resource availability mean that fuel cells are not selected for some weather years \cite{brown_neumann_etal_2025} (\cref{fig:capacities}(a); \cref{fig:fc_duration}).
However, a brief extreme event can upend system operations, spur investment, and recover installation costs for resilience back-ups.  
In particular, this impacts planning if such an event differs from the predominant weather conditions --- e.g. in years with a mild winter otherwise.

Resilience back-up capacities face financial risk due to inter-annual weather variability (\cref{fig:costs}): fuel cells bridge power deficits during system-defining events, and in the absence of said SDEs they remain unused.
This disproportionate reliance on a short-lived price spikes (Supplementary Figure S4) is more problematic due to low utilisation factors (\cref{fig:capacities}(g)), considering that a third of total electricity costs also falls into SDEs (\cref{fig:costs}(a)).
Other technologies, however, recover their capital cost over the entire span of the winter months (Supplementary Figure S5). 
Thus, the viability of fuel cells covering the power deficit in January 1966 is dubious in any other design year, shown by the large spread in annual revenue of resilience back-ups (though larger for fuel cells than gas power plants, \cref{fig:costs}(b)).

While the resilience back-up capacities as well as their cost recovery are highly sensitive, the identification of events and concrete challenges is much more robust (to assumptions such as transmission expansion, national self-sufficiency, and emission reductions; Supplementary Figure S6).
Regardless of the system configuration, we consistently identify SDEs characterised by large power deficits and requiring investment in resilience back-ups. 
On the one hand, considerations such as the choice of technology, national self-sufficiency, or transmission expansion are less relevant for resource adequacy during extreme events. 
On the other hand, the severity and costs of system stress can be slashed (Supplementary Figures S6--7) when CO$_2$ emissions are reduced by 99\% instead of 100\%, relative to 1990. 
In this setting, existing gas power plants cover peak generation entirely replacing fuel cells as resilience back-ups (\cref{fig:costs}(b), (c)) and operate similarly to fuel cells with low utilisation factors (Supplementary Figure S8).
However, the operation of existing gas power plants in this scenario is unprofitable and \emph{in none of the 80 years} are their capital costs as the resilience back-up recovered (\cref{fig:costs}(c)). 
The inclusion of other technologies (such as hydrogen turbines or gas power plants with CCS, see Methods) would lead to similar results due to analogous cost and utilisation fundamentals.
Through the set-up as a capacity expansion problem, all endogenous, newly installed infrastructure recovers its cost in the optimum (in a long-term market equilibrium), contrary to legacy infrastructure or (exogenously determined) retrofitting.
In particular, this elevates the importance of financial risk of providing sufficient back-ups to a European level, as the effects and system stress transcend national borders.

\subsection*{Variability and complexity of system-defining events}
\begin{figure}[H]
    \centering
    \includegraphics[width=\linewidth]{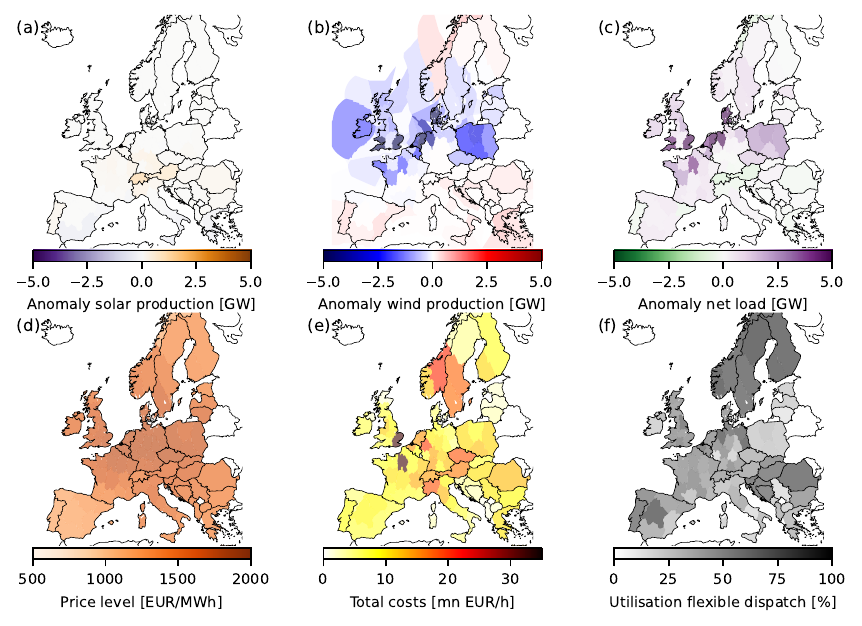}
    \caption{\textbf{Weather extremes propagate on a larger geographical scale than system impacts.} Composites of power system impacts during the identified SDEs. The maps show the averages across all 41 system-defining events for anomalies of (a) solar generation, (b) wind generation, (c) net load, as well as (d) the average shadow price, (e) average hourly system costs, and (f) the utilisation of flexibility capacities defined as in \cref{fig:capacities}(a). Maps made with \href{https://www.naturalearthdata.com/about/terms-of-use/}{Natural Earth}.}
    \label{fig:affected_areas}
\end{figure}

During system-defining events, the system copes by dispatching flexible generation according to marginal prices (ordered as in \cref{tab:characteristics} with the costs given in \cref{tab:cost_assumptions} in the Methods).
When wind droughts align with elevated electricity demand, this results in positive net load anomalies depicted in \cref{fig:affected_areas}(c). 
Through the high level of interconnection (Methods / Supplementary Figure S9), their impact on the system spans far beyond the low wind region: shadow prices spike continentally in \cref{fig:affected_areas}(d).
Simultaneously, flexible and resilience back-up generation is being activated on a system scale much larger than the area experiencing weather anomalies (\cref{fig:affected_areas}(a),(b),(f)), and the whole European system is covering net load as shown in \cref{fig:affected_areas}(c),(f).

\begin{figure}[H]
    \centering
    \includegraphics[width=\linewidth]{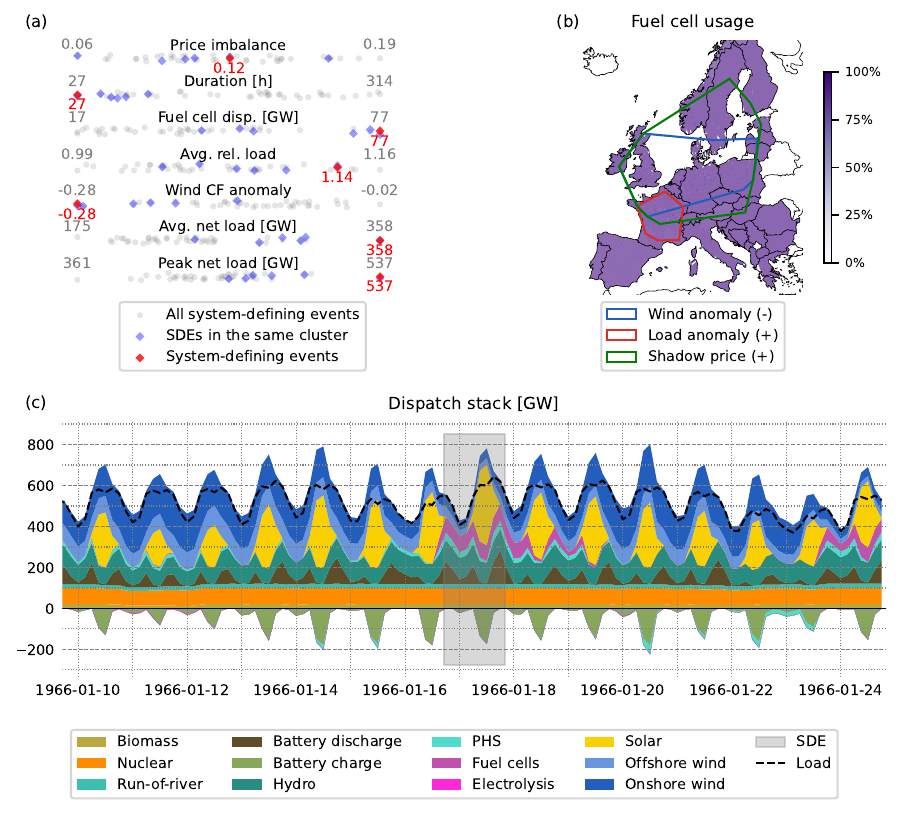}
    \caption{\textbf{Interactions between weather, network capacities, and economics during SDEs are complex.} Dashboard containing information about one identified severe power deficit event to exemplify characteristics of system-defining events. In (a), metrics assessing the severity of SDEs (Methods) show that event 11 (in January 1966) is the most extreme period identified when it comes to net load, wind anomaly, as well as fuel cell dispatch. The map in (b) shows that the areas affected by weather (load and wind anomalies) are smaller than the geographical extent of system effects (high shadow prices and fuel cell usage). (c) shows the dispatch stack during the SDE, as well as before and after, showcasing the renewable energy drought and how the flexibility capacities compensate it. Map made with \href{https://www.naturalearthdata.com/about/terms-of-use/}{Natural Earth}.}
    \label{fig:dashboard}
\end{figure}

Our filtering of SDEs based on shadow prices primarily captures power system stress, and we therefore investigate and cluster SDEs by their system impacts (Methods; \cref{fig:kpis-clusters}): \cref{fig:clusters}(b)-(e) reveals four types of events spanning from longer energy to shorter power capacity deficits (\cref{fig:clusters}(a)). 
The most extreme impacts are inflicted by \emph{severe power deficit events} which are relatively rare confluences of load and wind anomalies resulting in high net load (Supplementary Figures S10--13). 
We present the most extreme one in \cref{fig:dashboard}.
Although the short-lived price spikes recover capital cost in the optimisation model, the fuel cells are left unused for the remaining year (possibly the lifetime; \cref{fig:capacities}(g)) and might not provide resilience during other events (\cref{fig:fc_duration}).
SDEs in this cluster remain difficult and robust to the investigated sensitivities (Supplementary Figures S6--7). 
Another cluster consists of less extreme \emph{power deficit events} which rely less on fuel cells as a resilience back-up (Supplementary Figure S11).
Both occur regularly even though climate change signals over the 20th century tend to decrease winter heating load due to increasing temperatures (\cref{fig:net-load_trends}). 
If two power deficit events happen within a short time, they can form a \emph{cascading event} (Supplementary Figure S12). 
The effects of two, potentially different in nature, events compound and lead to prolonged system stress with high requirements for both energy and dispatch capacities. 
On the other extreme are \emph{energy deficit events} which are less consistently identified across different configurations (Supplementary Figures S6--7).
They carry a lower risk of stranded assets, and can be dealt with sufficient long-duration energy storage or allowing the use of fuel-based generators (using synthetic fuels or slightly relaxing CO$_2$ emissions).
As storage reserves are cheaper than fuel cells or other dispatch, additional storage capacities can serve as an investment in energy security (for purposes beyond weather resilience only), making energy deficits easier to plan for than power deficits which remain challenging even with the existing gas stock (Supplementary Figure S6).

\subsection*{Short- vs. long-term resilience challenges}
Short-lived power deficit events may be invisible in annual values (e.g. average heating load, annual capacity factors) which tend to correlate well with total system costs \cite{gotske_andresen_etal_2024}, a common measure to assess challenging weather years \cite{grochowicz_greevenbroek_etal_2023} (Supplementary Figure S14).
Low annual capacity factors influence capacities of renewables, whereas week-long SDEs rather drive back-up capacities, although not exclusively (\cref{fig:costs}(a)).

\begin{figure}[H]
    \centering
    \includegraphics[scale=1]{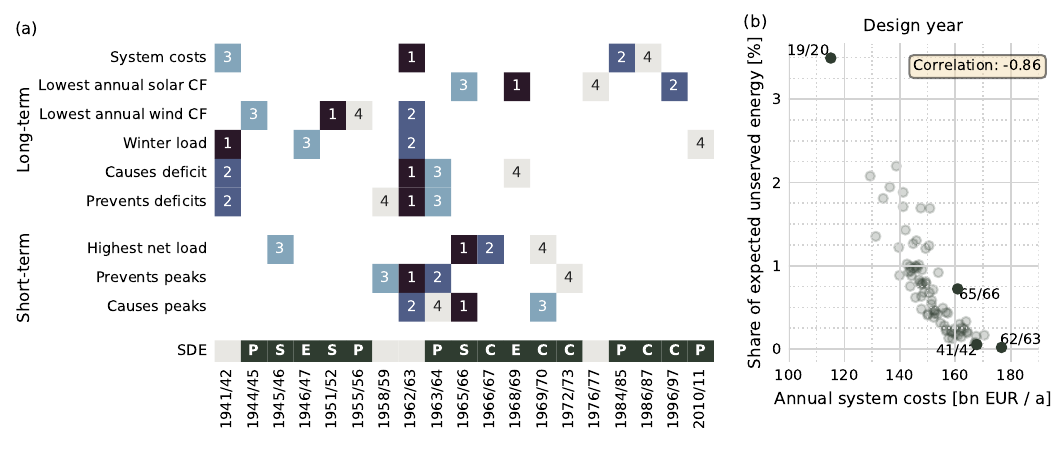}
    \caption{\textbf{Long-term reliability need not guarantee resilience against short-term stress during SDEs.} (a) shows the ranking of different weather years relating to metrics measuring resilience in short-term operation or long-term planning (the four most challenging weather years per metric; Methods). The long-term metrics are total system costs, annual capacity factors of wind and solar power, average electricity load during the winter, as well as expected energy not served (EENS) coming from the validation using dispatch optimisations. A weather year ``causes deficits'' if it being operated on other design years leads to high expected energy not served (EENS), and ``prevents deficits'' if its system layout leads to low EENS for other weather years. The short-term metrics show the years with the highest net load and the highest power deficits coming from the validation using dispatch optimisations. A weather year ``causes peaks'' if it being operated on other design years leads to the largest expected power deficit, and ``prevents peaks'' if its system layout leads to the lowest expected power deficit. We mark in green whether a year contains a system-defining event with the corresponding type of event (``S'', ``P'', ``C'', ``E'' correspond to severe power deficit, power deficit, cascading, and energy deficit events). Some challenging years such as 1941/2 contain no SDE. (b) shows the non-linear relationship and Pearson correlation between the total system costs based on a given weather year and the average share of expected unserved energy (EENS) when that capacity layout is operated on the remaining 79 weather years. 1962/3 and 1941/2 are highlighted as the years with the highest total system cost (long-term stress), 1965/6 for the highest net load (short-term stress), and 2019/20 for being the least adequate design year. See \cref{fig:years_stress_design} for an additional analysis.}
    \label{fig:short-long-term}
\end{figure}

For this reason, long-term and short-term resilience rely on different factors and ought to be measured in different metrics and tackled with different measures (\cref{tab:resilience_metrics}).
To identify drought years, high total system costs together with low renewable availability and high winter demand indicate years such as 1962/3, 1984/85, and 1941/42 (\cref{fig:short-long-term}) to be worth analysing. 
These metrics broadly agree with the computationally expensive validation (dispatch optimisation on all weather years; Methods) to identify the expected unserved energy (EENS; \cref{fig:years_max_median_reliability}; Supplementary Figures S15--19): a design with low annual EENS remains adequate under unexpected weather and an operational year causing high annual EENS is a good long-term stress test (\cref{fig:years_stress_design}).

On the other hand, short-term resilience is driven in periods on the order of days or weeks: extreme weather triggers system-wide back-up investments rapidly accruing shadow prices. 
Significant investment flows into resilience back-ups like fuel cells (sized by net load) which recover costs during these periods.
Again, validation on highest hourly unserved load (Methods) links identified SDEs with highest net load, but less so than for long-term metrics due to complex system-weather interactions.
Ultimately, design years with low maximal unserved load are more resilient to unexpected extreme weather; likewise, the extreme weather in 1965/66 is hard to operate and a suitable year with short-term stress tests.

Among the 80 considered weather years, no ``perfect storm'' of a harsh winter with an equally severe extreme event occurred.
The 1960s stand out as a decade of low wind resources \cite{wohland_omrani_etal_2019} and high load (Supplementary Figures S20--21) with 2010/11 being a notable, recent exception.
Weather resilience considerations neither need 80 weather years nor expensive validation ($80 \cdot 79$ optimisations; compare the quantitative analysis in Ref. \cite{furmann_gotske_etal_2025}), but should include long-term challenges (e.g. 1962/3) and years with short-term extremes as stress tests (e.g. 1965/66). 
As shown in \cref{fig:short-long-term} (b), handling extremes --- long- and short-term --- comes with a resilience premium and needs to trade-off energy security and energy affordability \cite{ruggles_virguez_etal_2024,schmitz_flachsbarth_etal_2025b} with a focus on the desirable resilience level.

\subsection*{Discussion}

Renewable energy droughts are an important concern in planning and operating net-zero power systems \cite{zhao_li_etal_2024}, and we show system-defining events (SDEs) can be responsible for a third of annual electricity costs.
While it is possible to design the European power \cite{grochowicz_greevenbroek_etal_2023} or energy system \cite{gotske_andresen_etal_2024} to withstand difficult weather years by increasing the costs by 10\% \cite{gotske_andresen_etal_2024}, the high electricity prices during periods we identify raise social concerns \cite{brown_neumann_etal_2025}.
Although a small admissible CO$_2$ budget (of up to 5\%) eliminates many SDEs, we find that existing dispatchable generation and daily balancing through batteries are not enough to cover the largest power deficits; rarely used resilience back-up capacities secure a reliable supply of electricity. 
Previous literature has discussed the dimensioning of the required long-term storage in highly renewable energy systems \cite{dowling_rinaldi_etal_2020, ruhnau_qvist_2022, brown_neumann_etal_2025, kittel_roth_etal_2026}. 
Optimal capacities for hydrogen storage vary by weather year (Supplementary Figure S3(b)), but its low unitary cost per energy capacity does not penalise installing storage overcapacity to improve resilience. 
Conversely, we see that the trade-off between adequacy and system cost is rather driven by the installed resilience back-up capacities.

Trading off energy security and affordability for net-zero power systems is complicated (\cref{fig:short-long-term}(b)) due to low utilisation factors of back-ups. 
Sufficient reserve capacities, fuel cells, CCS on fossil-based power plants, or hydrogen or methanol turbines (the latter three not included in the study) in net-zero scenarios or gas turbines in low-emission scenarios, compensate severe power deficits and provide resilience during extreme weather.
The exposure to financial risk of resilience back-ups --- recovering costs in reoccurring but irregular SDEs --- needs to be considered in planning to keep the capacities in reserve. 
Support schemes such as capacity mechanisms can address short-term stress, complemented by strategic reserves to manage long-term stress during challenging winters.

Our results align with ENTSO-E's European Resource Adequacy Assessment (ERAA) \cite{ERAA2024} which questions using existing assets for resilience until the end of their lifetime: 50~GW of new gas capacity would be needed by 2050, but low utilisation factors render much of existing gas power plants economically unviable by 2030 \cite{ERAA2024}.
Demand flexibility (e.g. from industrial consumers \cite{brown_neumann_etal_2025}) might reduce the need for resilience back-up, but to do so, intra-day flexibility will not be enough and a significant share of the demand needs to be deferred by a few days --- the length of many SDEs. 
Even with flexible hydrogen and synthetic fuel production, resilience back-up power generation capacities will be needed in some weather years \cite{gotske_andresen_etal_2024}.
Inter-annual variability drives system costs, the sizing of renewable capacities \cite{zeyringer_price_etal_2018} as well as back-up capacities \cite{gotske_andresen_etal_2024} and affects electricity price considerations \cite{brown_neumann_etal_2025}; and the focus of a study in question influences the number of necessary weather years \cite{ruggles_virguez_etal_2024}. 
For example, ENTSO-E's latest ERAA is based on three weather scenarios \cite{ERAA2024} minimising the distribution difference in revenues of thermal units, and disregards their meteorological difficulty or whether they caused system stress.
Though computationally expensive with 80 ($\cdot$ 79) optimisations, our approach systematically informs which weather years to model; studies as ours tied together with input weather screening\cite{furmann_gotske_etal_2025} can reduce the this to a small number of well-chosen weather years.
While we argue that dispatch optimisations are a superfluous computational burden, they can be useful in direct comparison of events across different weather years, even if design-dependent net load captures this well already.
Our results offer more nuance in distinguishing long-term resilience challenges from short-term ones: while total system costs capture long-term stress (e.g. 1962/3 or 1941/2), other metrics such as shadow prices and net load are more relevant for short-term stress (as in 1965/66). 
In particular, the critical periods are those with high power deficits which deviate from the predominant winter conditions. 
These most severe system-defining events do not have to overlap with challenging winters (high heating demand, low wind capacity factor) --- they did not in our dataset of 80 weather years --- and pose related, but distinct challenges to planning renewables-based power systems.

While our results suggest using a handful years combining short- and long-term resilience challenges (such as the event in January 1966 with the winter of 1962/3) instead of decadal time series, synthetic weather data or machine learning-based weather predictions may model hypothetical, extreme events\cite{price_sanchez-gonzalez_etal_2024}.
Using climate model outputs could quantify risks and return periods of unseen weather events or the impacts of climate change on critical weather conditions\cite{bloin-wibe_fischer_etal_2026}.
While historical data provides high-resolution stress tests, it may under-represent the frequency of future extreme events predicted by climate models. 
There is hope for higher spatial and temporal resolution outputs from CMIP7 for the climate impacts community following extensive discussion and consultation\cite{ruane_pascoe_etal_2025}.
Future work using these runs will allow for benchmarking of the historical stress tests in a warmer climate. 
Some more certain aspects of climate change are changing temperatures and precipitation patterns, which will lead to a shift in the timing of peak demands\cite{bloomfield_brayshaw_etal_2021} and reservoir dynamics \cite{vandermost_vanderwiel_etal_2024} (driven by changing drought frequencies), respectively. 
Even though climate change reduces heating demand in the winter, electrifying heating might increase the exposure to cold, still winter days \cite{bloomfield_2025}.
It is possible that decadal variations of wind generation (such as in the early 1960s where there were very cold and low wind periods present throughout the winters) may still occur even in a warmer climate \cite{vandermost_vanderwiel_etal_2025}. However, their likelihood is significantly reduced due to global warming.
While it remains to be seen whether frequency, severity or weather risks of SDEs will shift (results depend on the choice of climate model\cite{kapicaPotentialImpactClimate2024}), power deficit events will remain most challenging. 
A limitation in our optimisation set-up is the common assumption of perfect foresight throughout the year and the disregard of inter-annual energy storage.
However, for the short-term operation of batteries, current forecasting capacities validate this assumption, while for the long-term operation, Bellman values can be used to dispatch long-term storage similarly to a perfect-foresight scenario\cite{brown_neumann_etal_2025} and including hydropower storage limits based on historical records has shown to have negligible impact \cite{gotske_andresen_etal_2024}. 
While Gøtske et al. have assessed the robustness of the whole European energy system to weather variability, we choose to focus on the power system as it causes the largest resilience challenges also in a sector-coupled system \cite{gotske_andresen_etal_2024}. 
In particular, by keeping close to the framework from Grochowicz et al. \cite{grochowicz_greevenbroek_etal_2024}, we build on their analysis of weather conditions and focus on unanswered questions about system impacts.
Incorporating sector coupling, a wider array of resilience back-up technologies and more detailed carbon cycle tracking could clarify how much value each form of flexibility provides during SDEs.
Although (electrified) heating may impact the identified system-defining events, the inclusion of more energy sectors and carriers provides additional flexibility beyond the resilience back-up in our study (fuel cells or gas power plants).
Similarly, we disregard demand-side management which could provide even more flexibility \cite{brown_neumann_etal_2025} and resilience against extreme weather, possibly even rendering the values of shadow prices, which we use for identification of events and not as electricity prices, less degenerate.
Lastly, the desired levels of security can be weighed against the resilience premium, and concrete strategies for more resilient planning can be investigated.
It is likely that near-optimal solutions exist which exhibit less variability in back-up capacities; near-optimal spaces have been shown to be flat for renewable technology trade-offs \cite{pedersen_victoria_etal_2021,grochowicz_greevenbroek_etal_2023,vangreevenbroek_grochowicz_etal_2025} and might possess similar traits for flexibility capacities and their spatial allocation\cite{lombardi_vangreevenbroek_etal_2025}.
Our methods and analysis can be easily adapted to other studies and regions, and might lead to vastly different insights in systems relying on different generation mixes (or distinct transition pathways). 
Concrete challenges and timing of events might differ from the European situation, but the metrics and methods can identify and quantify both long- and short-term resilience needs.
A thorough understanding of specific weather triggers and system impacts during energy droughts can serve to design transformative strategies as well as evaluate market schemes that improve resilience into long-term energy system planning.

\section*{Methods}
\label{sec:methods}


\subsection*{Open model of the European power system}
We use the open-source energy system model PyPSA-Eur \cite{horsch_hofmann_etal_2018} (v0.13.0) which we restrict to the electricity sector in order to investigate the power system dynamics unaffected by other sectors.
PyPSA-Eur consists of a workflow, rooted in the PyPSA framework \cite{brown_horsch_etal_2018}, that conducts a partial greenfield optimisation based on existing energy infrastructure in Europe.
We assume existing (2023) generation capacities of gas, hydropower, nuclear, biomass, and (DC and AC) transmission, which are all fixed except for the latter. 
We use 2030 assumptions from \href{https://github.com/PyPSA/technology-data/releases}{technology-data v0.9.2} as in \cref{tab:cost_assumptions}. For transmission, we allow a 25\% expansion of current transmission volume (sum of lines length multiplied by capacity) compared to today's levels except for two cases that we explain in more detail below: (1) in the sensitivity scenarios assessing the impact of transmission expansion on SDEs (see Sensitivity analysis) and (2) in the validation through dispatch optimisations (see Validation).
For renewable generation (solar and wind) as well as energy storage through batteries and hydrogen, we assume a greenfield approach for lifetime reasons.
In contrast to Grochowicz et al.\cite{grochowicz_greevenbroek_etal_2024}, we disentangle charging and storage capacities of both hydrogen and batteries in order to better represent the trade-offs between additional energy and power capacities instead of a fixed power-to-energy ratio.
The model co-optimises the capacity expansion and dispatch to minimise annualised total system costs (formulated as a linear program) with a net-zero constraint on CO$_2$ emissions (except in a few sensitivity scenarios).
Our optimisations are based on highly resolved networks (90 nodes \cite{frysztacki_horsch_etal_2021}) of the European power system with hourly resolution for 80 different weather years (see below).

\subsection*{Flexibility}
In the set-up of this study, we have several options of providing flexibility to the supply side: on the one hand, transmission to move electricity across the network (expandable by 25\%, see above), and dispatch and storage to nodally inject additional generation capacities.
For the latter, we distinguish three different categories of flexible electricity supply:
\begin{enumerate}
    \item \emph{Existing dispatchable capacities} of nuclear power, biomass, and hydropower are assumed to be fixed and amount to approx. 300 GW. Although they have different operational properties, we lump them together as a fixed entity to assess whether the current stack can cover the needs of future net-zero systems.
    \item \emph{Daily balancing} is provided by (newly built) battery storage and is used consistently throughout the year, particularly in combination with solar generation. Their low energy-to-power ratio --- due to high energy capacity costs --- makes batteries more suitable for daily operations and not to cover renewable energy shortfall for multiple days (and less so in the winter).
    \item \emph{Resilience back-up} capacities are provided by (newly built) fuel cells (connected to hydrogen storage charged with electrolysers) in the default scenario. For our cost-optimisation approach, this is equivalent to substituting fuel cells by hydrogen or methanol\cite{glaum_neumann_etal_2026a}-fuelled gas turbines which have similar capital cost and efficiency. 
    
    These resilience back-ups are characterised by low utilisation throughout the year due to high marginal costs (low round-trip efficiency of hydrogen storage), and subject to substantial interannual variability in investments. In our sensitivity scenarios related to emission reductions, we include existing gas power plants (both OCGT and CCGT, anno 2023) that are seldom used outside extreme periods due to high marginal costs. The existing gas capacities are sufficient across all weather years. For this operational reason, we consider gas power plants as resilience back-ups and not as part of the existing dispatchable capacities. 
\end{enumerate}
We assume inelastic demand and perfect foresight for the entire year.

Furthermore, we exclude the option of adding carbon capture and storage (CCS) to existing (or new) gas power plants for computational reasons, as it would require tracking the carbon cycle. However, our analysis suggests the following:
\begin{itemize}
    \item A case in which CCS can be installed on existing gas power plants (or they can be retrofitted with H$_2$ turbines) would be structurally similar to the fuel cell case: these technologies would recover all of their costs (in contrast to the legacy gas stock), but still used at low capacity factors due to high marginal costs, even if slightly lower than fuel cells. Retrofitting a gas power plant with carbon capture adds a cost on the order of 600 EUR/kW \cite{rubin_zhai_2012,schroder_kunz_etal_2013}.
    \item A case in which H$_2$ turbines are allowed to be built from scratch will be structurally similar to the fuel cell case. The investment cost of H$_2$ turbines is lower than that of fuel cells (estimated at 550-1200 EUR/kW \cite{kost_mueller_etal_2024} which is higher than the OCGT value used for capital costs of H$_2$ turbines in PyPSA-Eur), but so is its conversion efficiency. Hence, newly-built H$_2$ turbines will perform similarly to fuel cells. In short, their investment cost is recovered by definition, but they operate for only a few hundred hours per year and not every year. 
    \item A case in which methane gas turbines with carbon capture can be installed from scratch is also similar to fuel cells since they will not emit CO$_2$ and will have similar cost (investment cost for OCGT is estimated at 460 EUR/kW, but adding carbon capture adds a cost of about 600 EUR/kW\cite{rubin_zhai_2012,schroder_kunz_etal_2013}, resulting in costs close to those assumed for fuel cell 1164 EUR/kW, and the latter attain higher efficiency).
    \item Finally, a case in which either carbon capture is exogenously added to all existing gas power plants, or all gas power plants are retrofitted to use hydrogen, will be similar to our case with gas power plants in the sensitivity analysis. In this case, the additional cost of adding the carbon capture or retrofitting will not be recovered, as the capacities are exogenously fixed. The resulting system will have zero CO$_2$ emissions, but the challenge of the gas backup plants not being able to recover their investment costs will persist.
\end{itemize}

\subsection*{Weather data}
We use 80 years of weather data from ERA5 \cite{hersbach_bell_etal_2020} (1941 -- 2021) that were translated into energy variables (capacity factors for solar generation, wind generation, and hydropower inflow) via atlite \cite{hofmann_hampp_etal_2021}. Electricity demand is adjusted for heating and cooling demand as in Ref.\cite{grochowicz_greevenbroek_etal_2023, vandermost_vanderwiel_etal_2022, frysztacki_vandermost_etal_2022} and also varies across the different weather years. We preserve winters and therefore run the model from July 1 to June 30 for each year\cite{grochowicz_greevenbroek_etal_2024,kittel_roth_etal_2026}, disregarding leap years. Compared to Section 2.1.1 of Grochowicz et al. \cite{grochowicz_greevenbroek_etal_2024}, we doubled the number of weather years, and added floating offshore wind and solar power with horizontal single-axis tracking which have since been included in PyPSA-Eur and are also processed with atlite.

\subsection*{Validation through dispatch optimisations}
For validation purposes, we also conduct dispatch optimisations fixing capacity layouts and varying the weather year as inputs (see below), thus measuring the adequacy of the optimised system design.
Though easier to solve than the capacity expansion problem, the validation through dispatch optimisation is a heavy computational burden (requiring $80 \cdot 79$ optimisations and 1.5 hours and 45 GB RAM per dispatch optimisaton resulting in a cumulative solving time just below 10 000 hours) offering few additional insights \cite{grochowicz_greevenbroek_etal_2024} except for \cref{fig:short-long-term}. 
In these runs, we allow for load shedding at a fixed cost of 100,000 EUR/MWh giving us a measure for unserved energy/load.
We fix transmission capacities to current levels for this analysis to avoid high load shedding and contaminated adequacy assessments due to different transmission investments across weather years.

\subsection*{Shadow prices and identification of extreme events/system-defining events}
We follow Grochowicz et al.\cite{grochowicz_greevenbroek_etal_2024} in their approach of identifying weather-driven periods of system stress as system-defining events (SDEs) during which a substantial amount of electricity costs is accumulated. Formally speaking, we look mostly at the shadow price $\lambda_{n,t}$ of the energy balance equation\cite{su_kern_etal_2020, schwaeppe_thams_etal_2024, brown_neumann_etal_2025, geis_neumann_etal_2026} 
\begin{equation}
\sum_g x^{g}_{n,t} + \sum_s x^{s}_{n,t} + \sum_{l} x^{l}_{t}   = d_{n,t}  \leftrightarrow \lambda_{n,t}
\end{equation}
at time $t$ in node $n$, where $x^{g}_{n,t}, x^{s}_{n,t}, x^{l}_{t}$ are the operational variable of generator $g$, storage $s$ at node $n$, line $l$ at time $t$, that is electricity generation, storage (dis-)charge and electricity import/export, respectively. The electricity load at time $t$ at node $n$ is denoted by $d_{n,t}$.
The shadow price is a good indicator of system stress, as it connects demand, (renewable) generation availability with the network layout and storage levels.
A period is considered a system-defining event if it accumulates electricity costs (system-wide sum of shadow priced weighted by load) beyond a given threshold (here, $C = 100$ bn EUR and $T = 14$ days; this accounts to 35\% of annual electricity costs on average):
\begin{equation}
    \sum_n \sum_{t=t_0}^{t_0 + T - 1} d_{n,t} \cdot \lambda_{n,t} \geq C,
\end{equation}
These thresholds yielded on average one SDE every 1.5 years \cite{grochowicz_greevenbroek_etal_2024}; our improved representation of storage decoupling energy and power capacities reduced the frequency of SDEs to once every two years.

It should be remarked that our filtering with shadow prices is skewed towards uneven conditions, i.e. if some weather period stands out from the remaining winter (or weather year), it is more likely to be identified, as it triggers additional and distinct investment. 
Thus, a few SDEs are not as ``extreme'' as those that are consistently identified across sensitivity scenarios, for instance.
We also emphasise larger geographical scales through the lens of costs (weighted by demand); hence in theory small differences in system layout, system variables, or just the existence of another SDE during the same year can impact the classification into system-defining or not.
An additional analysis of periods whose weather conditions are worse than some SDEs shows no larger challenges to the system. 
The robustness of identified events and their ranking across scenarios strongly suggests we have filtered out the most extreme events for the system, even if we have no guarantee to have found the ``most extreme'' weather. 

For a more detailed description of different shadow prices and how the duration of a system-defining event is determined we refer to the supplementary material of Ref.\cite{grochowicz_greevenbroek_etal_2024}. 

Although net load defined as the difference of demand and renewable generation,
\begin{equation}
    \text{net load}_t = \sum_n d_{n,t} - (\sum_n x^{\text{solar}}_{n,t} + x^{\text{wind}}_{n,t}),
\end{equation}
can be used to determine severity of an extreme event, it is less suitable as a mode to identify extreme events. Similarly, load shedding in a dispatch optimisation does not improve the identification of extreme events and has other significant disadvantages (computational burden, required capacity layout). Despite this, we use both to further assess and strengthen the robustness of our results; in particular net load can complement the weakness of a shadow price-based identification which cannot compare the severity of events across different optimisations (that is, across weather years).

In this study, we summarise the characteristics of system-defining events in Europe in \cref{tab:characteristics}.

\subsection*{Clustering of events}
Due to the difficulty of clustering all factors contributing to system stress during system-defining events (variables pertaining to renewable availabilities, transmission and storage operation resolved at every hour and node), we use aggregated system metrics that we deemed most insightful after an initial analysis to cluster SDEs.
The seven metrics we consider are
\begin{itemize}
    \item highest net load [GW],
    \item average net load [GW],
    \item duration [hours],
    \item total fuel cell discharge [TWh],
    \item maximal fuel cell discharge [GW],
    \item average relative load [unitless],
    \item wind capacity factor anomaly [unitless],
\end{itemize}
which we all normalise during the cluster initialisation.
According to Silhouette scores and Calinski-Harabasz score \cite{schubert_2023}, we choose four clusters to represent the system-defining events (see \cref{tab:clusters}).
We disregard a more granular geographical representation of weather and system variables, as we did not see any additional merit during an initial analysis. For single days of SDEs (also across distinct SDEs) a more granular geographical clustering of weather variables has already been conducted \cite{grochowicz_greevenbroek_etal_2024}. Additionally, the relationship between SDE and the encompassing weather year appears complex, which we investigated with Wasserstein metrics assessing the similarity of weather years and their SDEs (through net load and wind capacity factors; Supplementary Figures S22--23).

\subsection*{Resilience metrics}
We assess the system adequacy and resilience of the capacity layouts that we obtain and analyse in this study with the  metrics presented in \cref{tab:resilience_metrics}.
Related to the definitions by Schmitz et al. \cite{schmitz_flachsbarth_etal_2025b}, most of our metrics relate to system adequacy (capacity to satisfy energy requirements while complying with requirements on security and quality of service), resilience (ability to retain, react, overcome, and overpass perturbations caused by a shock), and reliability (probability of satisfying the load demand under uncertain conditions). For more details, see \cref{tab:resilience_metrics}.

The seven metrics we use to cluster the events are closely related to the resilience metrics in \cref{tab:resilience_metrics} (with net load and dispatch capacities included, as well as temperature and wind resources on a shorter time scale). As resilience is driven by the system (layout) and relative to adequacy requirements, measuring both severity and resilience is case-dependent.

\subsection*{Sensitivity analysis}
To assess the robustness of the identified system-defining events across different system configurations, we conduct a sensitivity analysis rooted in three key assumptions. 
We vary the transmission levels, the assumed reduction in CO$_2$ emissions as well as requirements to generate enough electricity nationally across the different countries in Europe.
Beside the default assumptions of an admissible transmission volume expansion of 25\% on top of the existing interconnections in Europe, we also explore scenarios prohibiting any additional transmission expansion as well as runs which allow doubling of transmission volume (in TWkm).
Moreover, we explore the impact of enforcing self-sufficiency requirements (by default there are none) in the European power system where every country needs to cover 70 or 90\% of its annual electricity demand nationally.
Lastly, we investigate the effect of less ambitious CO$_2$ emissions reduction targets, 95 and 99\%, relative to 1990, instead of 100\% as in the default net-zero scenario. 
With the option to emit a limited amount of CO$_2$, existing gas power plants can be used to supply electricity (up to 2.5\% of annual generation for a 99\% emission reduction), replacing all hydrogen storage and fuel cells in the cost-optimal solutions.
The usage of existing gas power plants (assumed to be amortised) instead of adding hydrogen infrastructure reduces shadow prices and thus the triggered investments during system-defining events.
For this reason, we vary the threshold used in the identification of SDEs; this helps to answer if different system configurations dampen the impact of extreme weather during SDEs and whether the same types of SDEs are identified at lower thresholds (and thus lower intensity).

Although we looked at the combination of all of the above scenarios, resulting in 27 different sensitivity runs, we only show the one-at-a-time sensitivity in Supplementary Figure S6. 
The remaining sensitivities do not add new insights, as the impact of transmission expansion and national self-sufficiency requirements both on the number of identified system-defining events (Supplementary Figure S7) and robustness of the SDEs in the default scenarios is negligible.

\section*{Data availability}

All data used are open (various licenses). Intermediate data can be found at \url{https://zenodo.org/records/20590389} \cite{dataset_grochowicz_2025}.

\section*{Code availability}

All code is open source (licensed under GPL v3.0 and MIT). The code and instructions to reproduce the study can be found in the following GitHub repository: \url{https://github.com/aleks-g/stressed-system/tree/v1.0} (with a copy in \cite{dataset_grochowicz_2025}).

\section*{Correspondence}
Correspondence and requests for material should be addressed to Aleksander Grochowicz (\texttt{\url{algro@dtu.dk}}).

\section*{Acknowledgements}

ERA5 reanalysis data \cite{hersbach_bell_etal_2020} were downloaded from the Copernicus Climate Change Service (C3S) \cite{c3s-2023}. The results contain modified Copernicus Climate Change Service information 2020. Neither the European Commission nor ECMWF is responsible for any use that may be made of the Copernicus information or data it contains.

The authors gratefully acknowledge the computational and data resources provided on the Sophia HPC Cluster at the Technical University of Denmark \cite{SOPHIA}.

\section*{Funding Statement}

A.G. and M.V. acknowledge funding from DFF Sapere Aude—EXTREMES project (2067-00009B). H.C.B. acknowledges funding from a Newcastle Academic Track Fellowship.

\section*{Author Contributions Statement}

Conceptualisation: all authors. Methodology: A.G. Coding and analysis: A.G. Writing - original draft: A.G. Writing - review and editing: all authors. Supervision: M.V. Funding: M.V.

\section*{Competing Interests Statement}

The authors declare no competing interests.

\section*{Extended Data}
\setcounter{figure}{0}
\renewcommand{\thefigure}{E\arabic{figure}}
\renewcommand{\thetable}{E\arabic{table}}

\begin{figure}[H]
    \centering
    \includegraphics[scale=1]{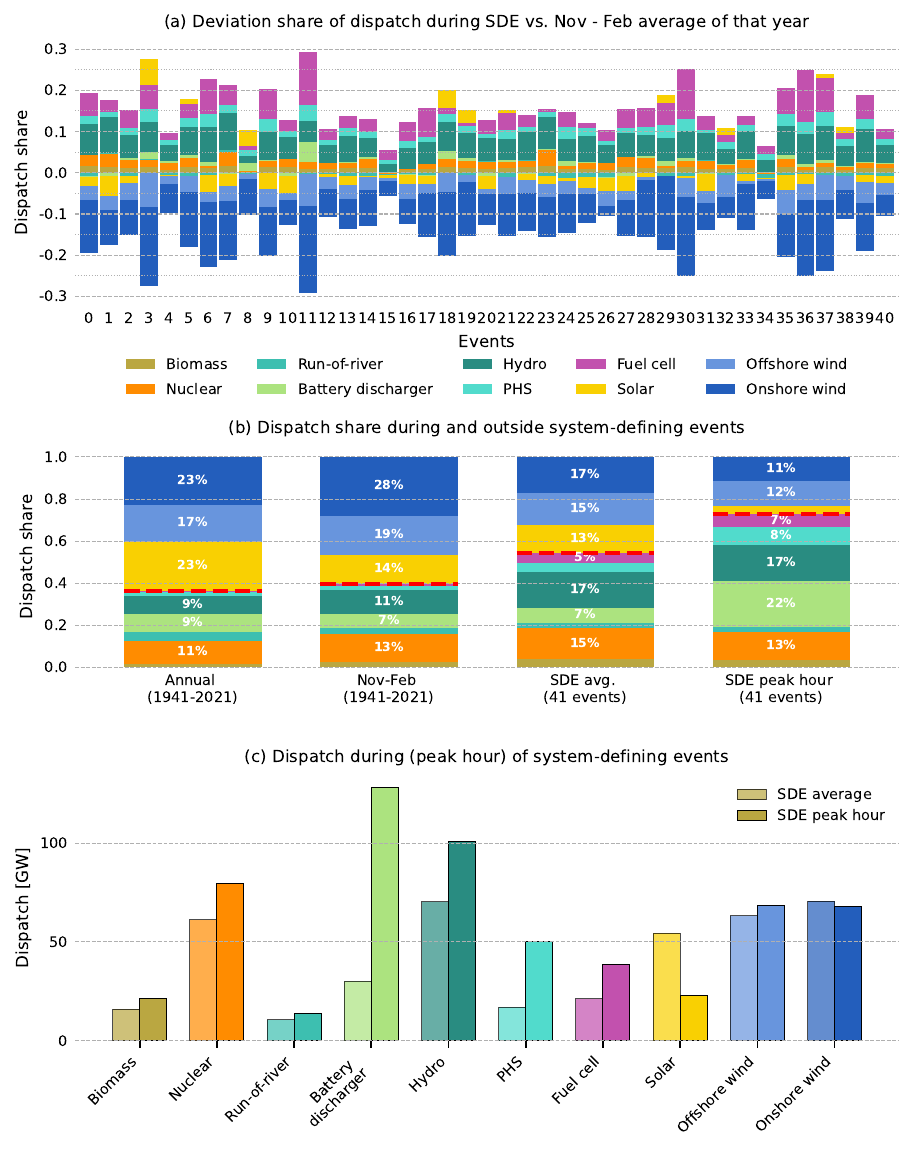}
    \caption{\textbf{Hydropower and fuel cells compensate the dispatch deficit from wind droughts}. (a) shows the difference in dispatch by technology between each system-defining event and the average share between November and February. (b) compares the dispatch share of different technology at different time scales. (c) compares the dispatch of different technologies between the peak hours and average of SDEs. Note that the spikes in battery discharge during peak hours (usually the evening peak) are not a feature of SDEs, but of the daily charging/discharging cycle.}
    \label{fig:combined_gen_analysis}
\end{figure}

\begin{figure}[H]
    \centering
    \includegraphics[scale=1]{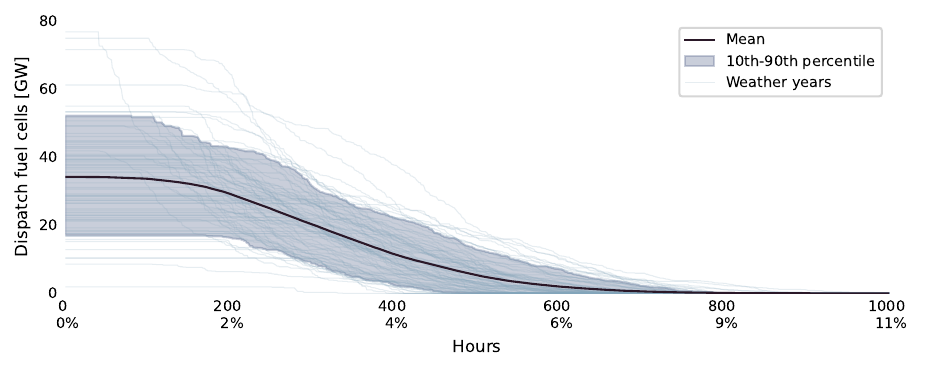}
    \caption{\textbf{Fuel cells are utilised less than 10\% of the year and rarely at full capacity}. Duration curve of fuel cell discharge. Different weather years (80 in total) are indicated as well as the 10th and 90th percentile ranges. The large variations stem from different installed fuel cell capacities that would either be unviable or not needed in other weather years.}
    \label{fig:fc_duration}
\end{figure}

\begin{table}[H]
    \centering
     \begin{tabular}{@{}p{3.5cm}p{1.7cm}p{1.7cm}p{2.5cm}p{1.5cm}p{2.5cm}p{2cm}@{}}
        \toprule
        \textbf{Cluster} & \textbf{Severity} & \textbf{Number of events} & \textbf{Return period (approx.)} & \textbf{Duration} & \textbf{Resilience back-up needs} & \textbf{Storage usage} \\ \midrule
        \textbf{Severe power deficits} & Very high & 6 events & 13 years & 1--4 days & Very high & Moderate \\
        \textbf{Power deficits} & Moderate & 12 events & 7 years & 2--8 days & Moderate & Low \\
        \textbf{Cascading events} & High & 9 events & 9 years & 7--12 days & High & Very high \\
        \textbf{Energy deficits} & Low & 14 events & 6 years & 7--14 days & Low & High \\
        \bottomrule
    \end{tabular}
    \caption{Identified clusters with key characteristics; see \cref{fig:kpis-clusters} for more details.}
    \label{tab:clusters}
\end{table}

\begin{table}[H]
    \centering
    \begin{tabular}{p{4cm}p{14cm}}
    \toprule
    \textbf{Property} & \textbf{Description} \\ \midrule
        \textbf{Time of occurrence} & November through February, higher probability in December and January  \\
        \textbf{Return times} & approx. every 2 years \\
        \textbf{Meteorology \& Renewables} & Wind drives the occurrence of extreme events, but (lack of) solar due to seasonality is the necessary condition; high heating demand \\
        \textbf{Geographical scale} & varying; limited for weather impacts; system impacts (prices / flexibility usage) larger if not continental \\
        \textbf{Duration} & 1 day -- 10 days (avg. 1 week) \\
        \textbf{Severity} & varying, system-dependent, many criteria (see later) \\
        \textbf{Identification} & Shadow prices, threshold on accumulated electricity cost \\
        \textbf{Challenge} & Provision of sufficient power / flexible generation, secondary energy provision \\
        \textbf{Order of dispatch} & Renewables, battery, nuclear, biomass, hydropower, pumped hydro storage, fuel cells \\ 
      \bottomrule
    \end{tabular}
    \caption{Characterisation of the system-defining events (SDEs) identified in this study.}
    \label{tab:characteristics}
\end{table}

\begin{figure}[H]
    \centering
    \includegraphics[scale=1]{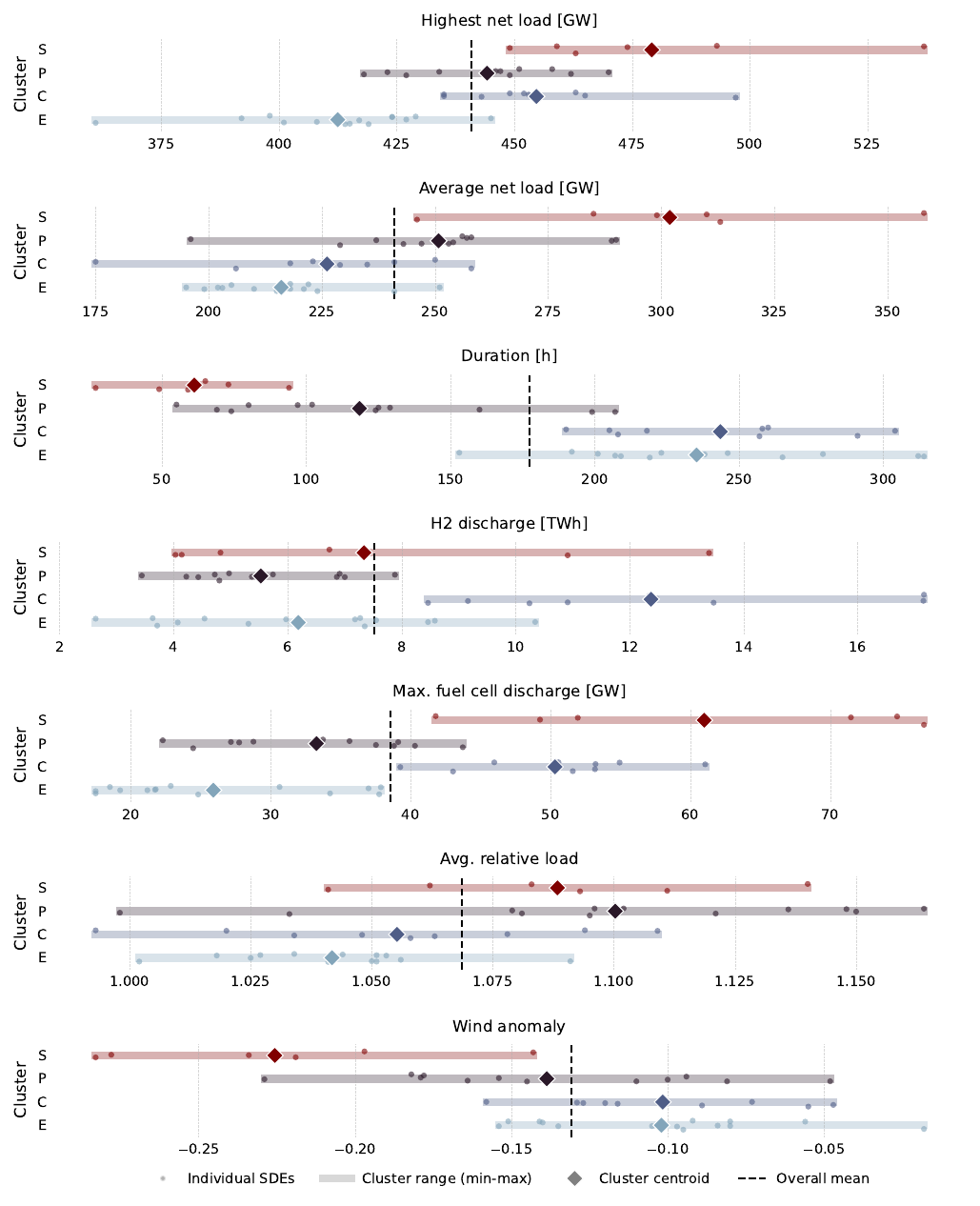}
    \caption{\textbf{SDEs can be clustered along a spectrum from extreme power deficit to energy deficit events}. Performance of the different SDE, sorted by cluster, according to the different metrics used for clustering. The means for each metric are marked with a dashed line as well as the centroid and range between mix and max of each cluster. ``S'', ``P'', ``C'', ``E'' stand for the four cluster names (severe power deficit event, power deficit event, cascading event, energy deficit event) respectively. Each point corresponds to one of 41 SDEs.}
    \label{fig:kpis-clusters}
\end{figure}

\begin{figure}[H]
    \centering
    \includegraphics[scale=1]{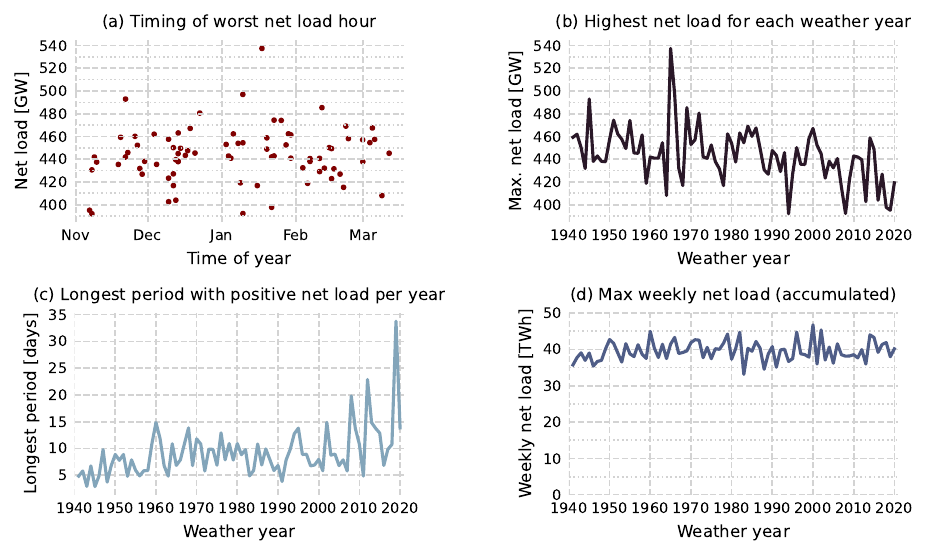}
    \caption{\textbf{Climate change tends to decrease maximal annual net load due to reduced heating demand}. Trends for net load metrics across different weather years suggesting signals from climate change: (a) timing of the worst hour during the year (for each of the 80 weather years), (b) maximal net load, (c) longest period with positive net load (i.e. more cumulative demand than renewable generation, disregarding severity), (d) maximal weekly net load. }
    \label{fig:net-load_trends}
\end{figure}

  \begin{table}[H]
      \centering
      \begin{tabular}{@{}p{3.9cm}p{2.9cm}p{4cm}p{1.6cm}p{1.4cm}p{2.2cm}@{}}
          \toprule
          \textbf{Technology} & \textbf{Capital cost} & \textbf{Marginal cost} & \textbf{Efficiency} & \textbf{Lifetime} & \textbf{Expandable} \\
          \midrule
          \multicolumn{6}{c}{\textit{Generation}} \\
          Solar & 543 EUR/kW$_e$ & -- & cap. factor & 40 years & yes, greenfield \\
          Solar, single-axis tracking & 629 EUR/kW$_e$ & -- & cap. factor & 40 years & yes, greenfield \\
          Onshore wind & 1096 EUR/kW & -- & cap. factor & 30 years & yes, greenfield \\
          Offshore wind & 1682 EUR/kW$_e$ & -- & cap. factor & 30 years & yes, greenfield \\
          Offshore wind, floating & 2350 EUR/kW$_e$ & -- & cap. factor & 20 years & yes, greenfield \\
          Nuclear & -- & 3.55 EUR/MWh (incl. fuel) & varying & 40 years & no \\
         Biomass & -- & 7.4 EUR/MWh$_{th}$ & 47\% & 30 years & no \\
          Closed-cycle gas turbine$^*$ & 878 EUR/kW & 4.44 EUR/MWh & 58\% & 25 years & no$^*$ \\
          Open-cycle gas turbine$^*$ & 461 EUR/kW & 4.76 EUR/MWh & 41\% & 25 years & no$^*$ \\  
          \midrule
          \multicolumn{6}{c}{\textit{Storage}} \\
          Battery storage & 150 EUR/kWh & -- & -- & 25 years & yes, greenfield \\
          Battery inverter & 169 EUR/kW & -- & 96\% & 10 years & yes, greenfield \\
          H$_2$ underground storage & 2 EUR/kWh & -- & -- & 100 years & yes, greenfield \\
          Fuel cell & 1164 EUR/kW$_e$ & endogenous & 50\% & 10 years & yes, greenfield \\
          Electrolyser & 1500 EUR/kW$_e$ & -- & 62\% & 25 years & yes, greenfield \\
          \midrule
          \multicolumn{6}{c}{\textit{Transmission}} \\
          HVDC converter pair & 165,803 EUR/MW & -- & & 40 years & yes, limited \\
          HVDC/HVAC overhead & 442 EUR/MW/km & -- & & 40 years & yes, limited \\
          HVDC underground & 1008 EUR/MW/km & -- & & 40 years & yes, limited \\
          \midrule
          \multicolumn{6}{c}{\textit{Fuel costs}} \\
          Natural gas$^*$ & -- & 24.5 EUR/MWh$_{th}$ & -- & -- & -- \\
          \bottomrule
      \end{tabular}
      \caption{Cost assumptions for key technologies. A discount rate of 7\% is assumed for cost annualisation. Technologies available in the sensitivity analysis are marked with a star. Hydropower, and pumped hydro storage are assumed to be fixed (with no fuel costs) and are thus not included in this table. The values are 2030 assumptions from \href{https://github.com/PyPSA/technology-data/releases}{technology-data v0.9.2}}
      \label{tab:cost_assumptions}
  \end{table}

\begin{table}[H]
    \centering  
    \begin{tabular}{@{}p{2.8cm}p{1.7cm}p{1.7cm}p{3cm}p{5cm}p{3cm}@{}}
        \toprule
        \textbf{Metric} & \textbf{Horizon} & \textbf{Aspect} & \textbf{Definition} & \textbf{Related Question} & \textbf{Use-case} \\ \midrule
        Shadow prices & Short-term, hours/days & Reliability / Resilience & see system-defining events (SDEs) & How much were capacity investments, both renewable and flexibility, affected and triggered in this period? & Identification of stress \\
        Net load & Short-term, hours/days & Adequacy / Reliability & Difference of demand and renewable generation & How much flexible capacity is needed in the system? & Severity of stress \\
        Prevents peaks & Short-term, hours & Reliability & Max. unserved energy for design year & Is the system resilient to other extremes? Does the extreme event brace us for other events? & Planning systems resilient to short-term stress \\
        Causes peaks & Short-term, hours & Reliability & Max. unserved energy as operational year & How difficult is the period in other systems? Is it a good stress test for extremes? & Stress-testing systems \\
        Dispatch capacities & Short-term, long-term & Adequacy & Capacities of flexible dispatch (e.g. fuel cells, battery dischargers) & What infrastructure (size) is needed? & Comparing difficulty of weather years and short-term stress \\
        System costs & Long-term, annual & Adequacy & Total system costs & What infrastructure (size) is needed? & Comparing difficulty of weather years \\
        Solar resources & Long-term, seasonal & Adequacy & Annual solar capacity factor & What infrastructure (size and mix) is needed? & Comparing difficulty of weather years \\
        Wind resources & Long-term, seasonal & Adequacy & Annual wind capacity factor & What infrastructure (size and mix) is needed? & Comparing difficulty of weather years \\
        Temperature & Long-term, seasonal & Adequacy & Average electricity load during winter months & What infrastructure (size) is needed? & Comparing difficulty of weather years \\
        Prevents deficits & Long-term, annual & Resilience & Expected energy not served (EENS) for design year & How much unserved energy does this system have with unexpected weather? Is it resilient to weather variability? & Planning systems resilient to long-term stress \\
        Causes deficit & Long-term, annual & Resilience & Expected energy not served (EENS) as operational year & How challenging is this weather year? Is it a good stress test? & Stress-testing systems \\
        \bottomrule
    \end{tabular}
    \caption{Resilience metrics}
    \label{tab:resilience_metrics}
\end{table}

\begin{figure}[H]
    \centering
    \includegraphics[scale=1]{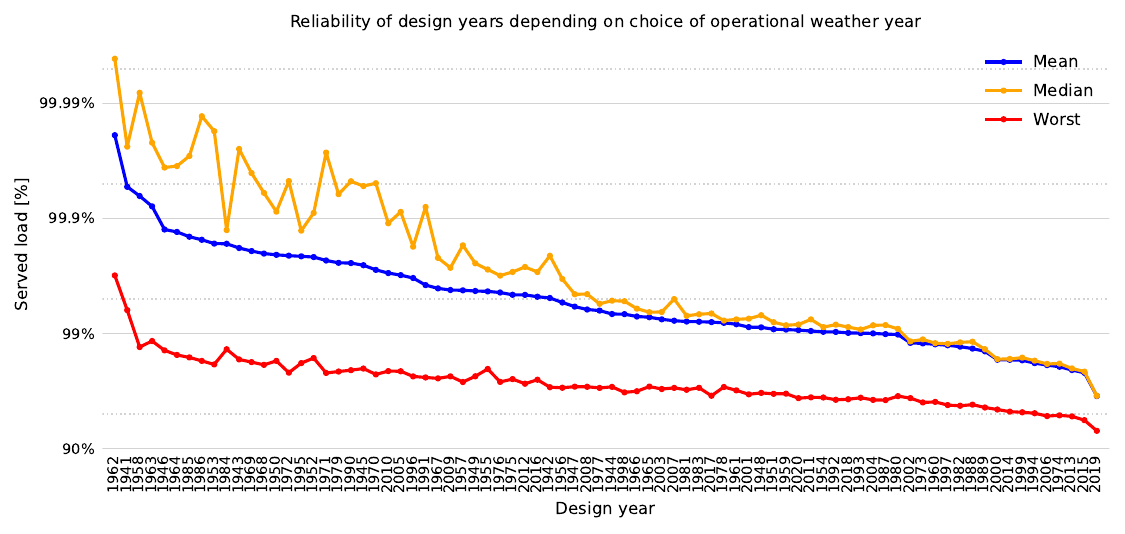}
    \caption{\textbf{Only 2/80 years could have served more than 99\% of load across the investigated weather years}. Reliability of system designs across weather years. Mean and median values for served load are marked as well as the value for the worst operational year. Note the log-scale on the y-axis.}
    \label{fig:years_max_median_reliability}
\end{figure}

\begin{figure}[H]
    \centering
    \includegraphics[scale=1]{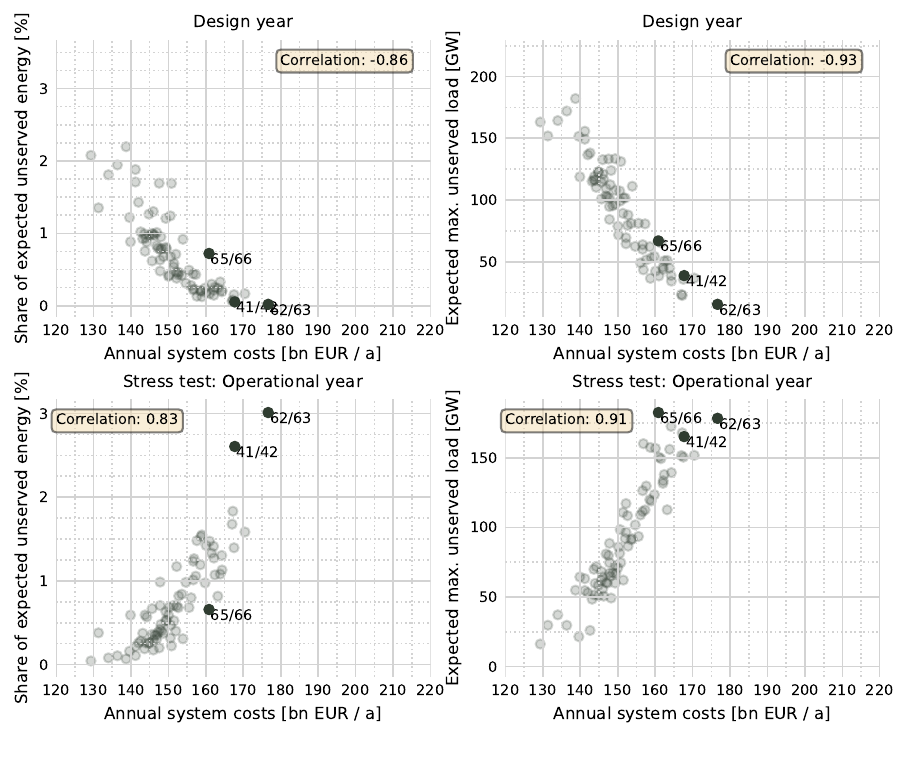}
    \caption{\textbf{Some years are better stress testers for long-term resilience (e.g. 62/63 for annual EENS) than for short-term resilience (e.g. 65/66 for maximal unserved load)}. Correlation between total system costs of design years and reliability as design and operational years (for expected energy not served (EENS) and maximal unserved load). (a) is the same as \cref{fig:short-long-term}(b); 1962/3 and 1941/2 are marked as the years with the highest total system cost (long-term stress), and 1965/6 for the highest net load (short-term stress). Each dot represents one out of 80 weather years and the correlation describes the Pearson correlation coefficient.}
    \label{fig:years_stress_design}
\end{figure}

\bibliographystyle{naturemag}
\bibliography{bibliography_final.bib}

\end{document}


\section*{Supplementary information for ``Resilience metrics to guide back-up investments in the power system during extreme weather''}

\subsection*{Dispatch shares during winter months and system-defining events (SDEs)}
\begin{figure}[H]
    \centering
    \includegraphics[scale=1]{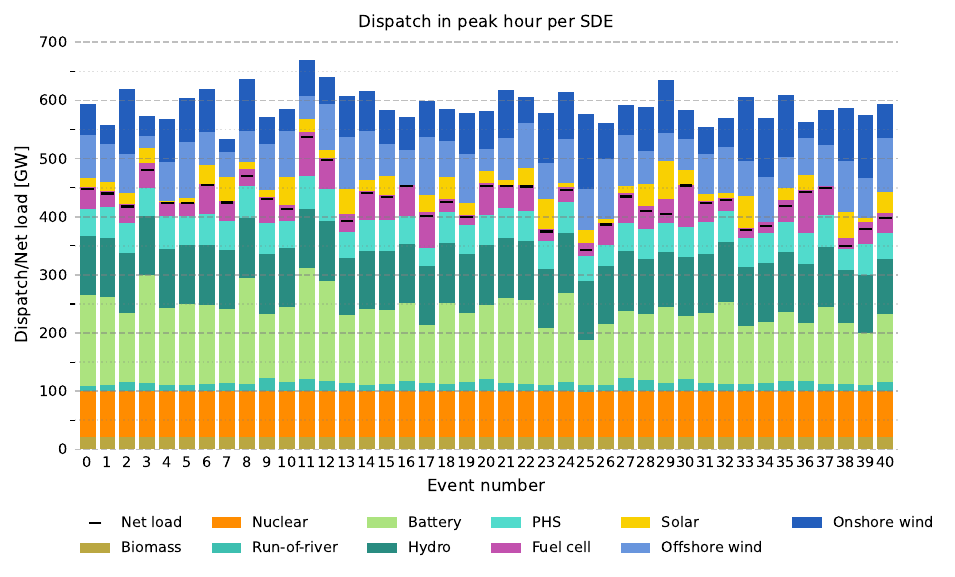}
    \caption{\textbf{The severity of system-defining events (SDEs) during their worst hour differs significantly across events}. Dispatch and net load during the peak hour (defined as the hour with the maximal net load) of each system-defining event.}
    \label{fig:sde_peak_gen_net_load}
\end{figure}


\subsection*{Utilisation and capacities of resilience back-ups and storage}

\begin{figure}[H]
    \centering
    \includegraphics[scale=1]{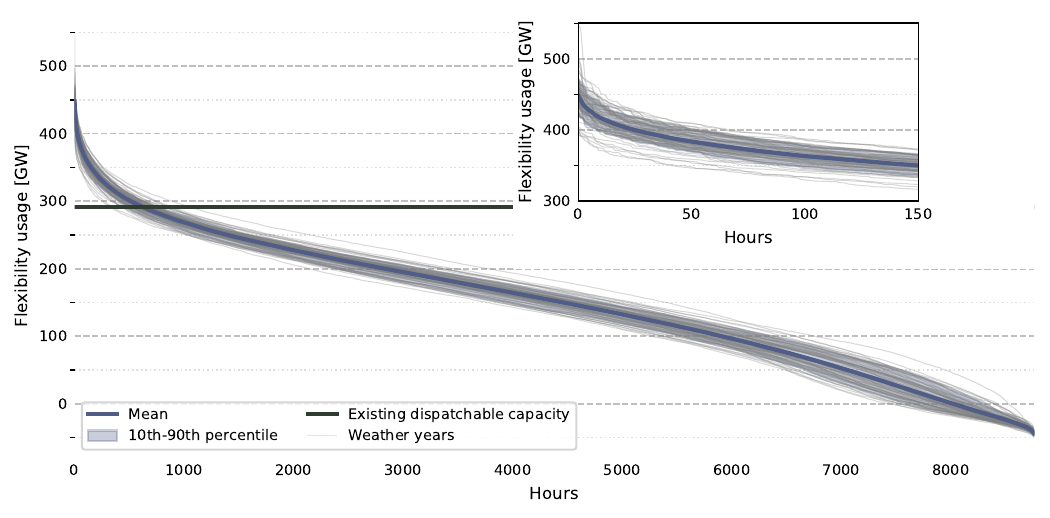}
    \caption{\textbf{The need for resilience back-ups is relatively short and highly variable across weather years}. We show the duration curve of flexibility usage, accumulated for the three types of flexible electricity supply (existing dispatchable capacities, daily balancing, resilience back-ups; see Methods and Figure 3 in the main paper). We zoomed in on the first 150 hours where fuel cells are used heavily. Different weather years (80 in total) are indicated as well as the 10th and 90th percentile ranges.}
    \label{fig:net_load_duration}
\end{figure}


\begin{figure}[H]
    \centering
    \includegraphics[width=\textwidth]{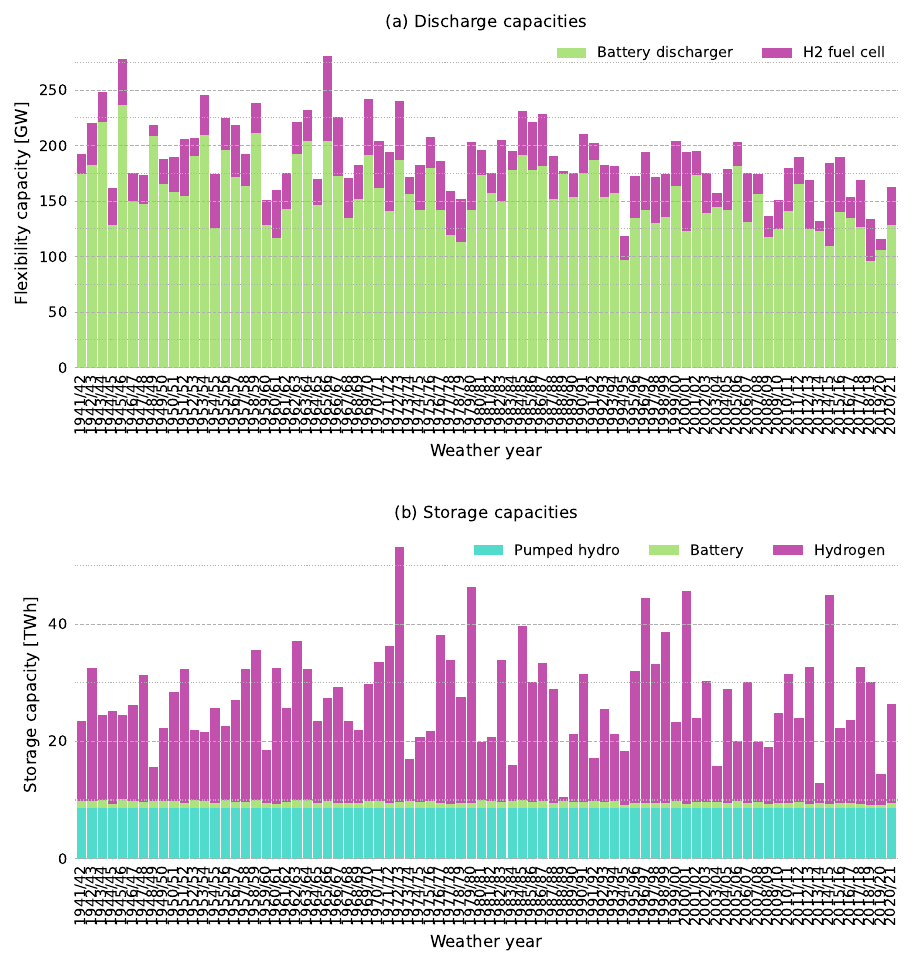}
    \caption{\textbf{Power/Discharge capacities of storage is much more variable than storage capacity.}. (a) Annual capacities of battery discharge and H2 fuel cells. The existing capacities of nuclear, hydropower, and biomass (see Figure 3 in the main paper) are not shown as they are fixed in the optimisations. 1945/46 and 1965/66 have the highest capacities for battery dischargers (236 GW) and fuel cells (77 GW), respectively. (b) Annual storage capacities of different weather years; capacities for pumped hydro storage are fixed. Although valuable during extreme events, the storage capacity (175 TWh) of hydro reservoirs is not included, as it depends on inflow, and is operated as a flexible dispatch in contrast to the other technologies shown.}
    \label{fig:flex_caps}
\end{figure}

\subsection*{Annual cost recovery and hourly electricity costs}
\begin{figure}[H]
    \centering
    \includegraphics[scale=1]{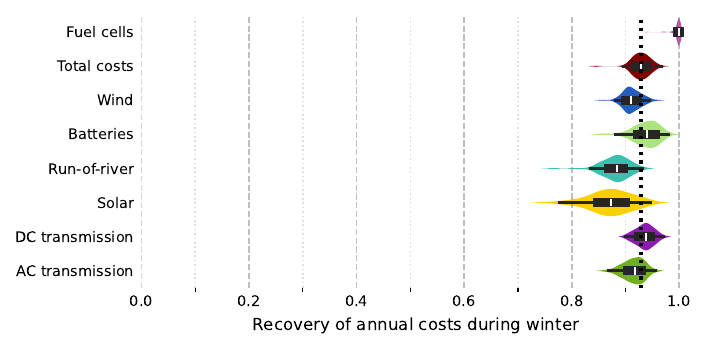}
    \caption{\textbf{All system component recover almost all of their cost only during the winter months}. Recovery of different components during the winter months relative to annual costs. Similar to Figure 1(a). The whiskers show the minimum and maximum values, whereas the boxes show the median (in white) and the lower and upper quartiles for 80 weather years.}
    \label{fig:recovery_winter}
\end{figure}

\begin{figure}[H]
    \centering
    \includegraphics[scale=1]{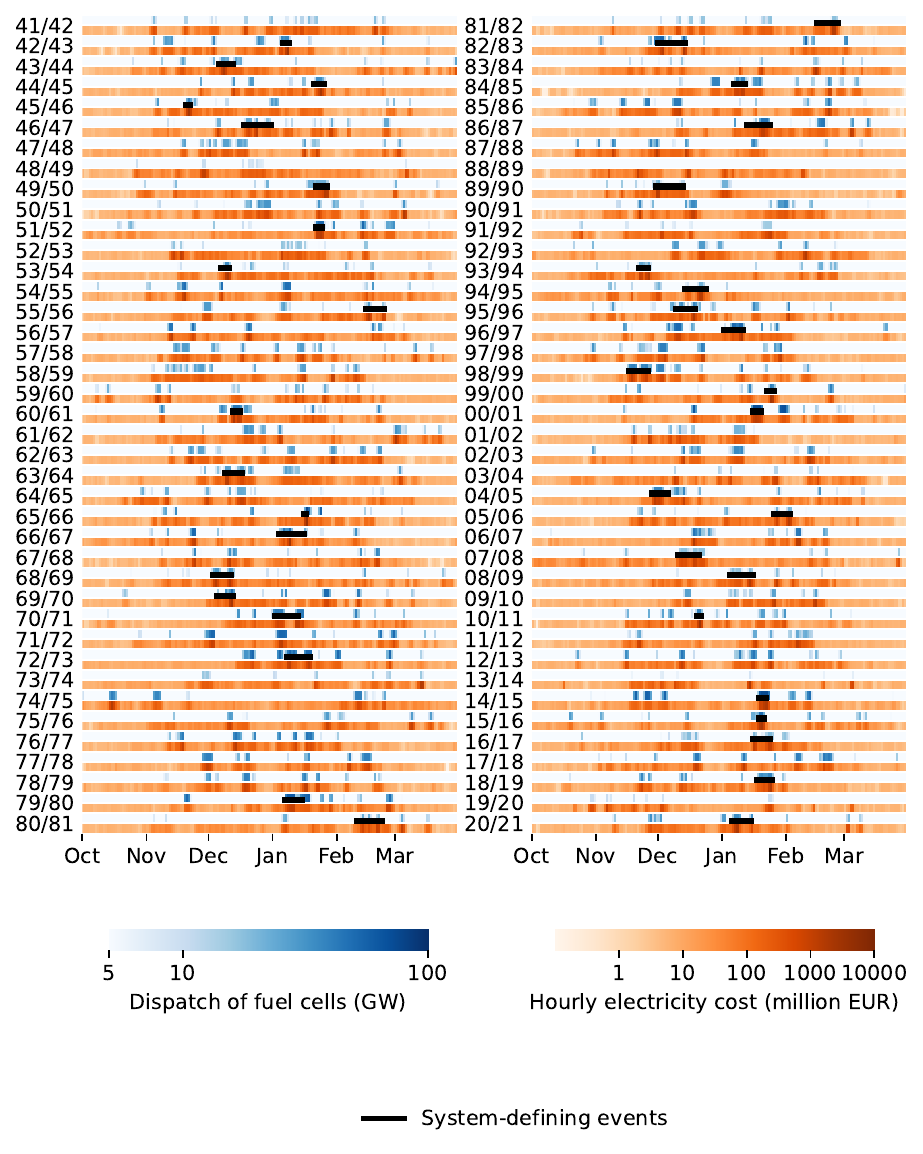}
    \caption{\textbf{Hourly electricity costs and fuel cell usage spike in the winter months and coincide well}. Hourly electricity costs in the different optimal networks and utilisation of fuel cells, with system-defining events marked. Only the period between October and March is shown, as hourly electricity costs in the summer are low, and fuel cells are barely utilised.}
    \label{fig:dual_costs_sde}
\end{figure}

\newpage
\subsection*{Sensitivity analysis and behaviour of gas as a resilience back-up}

\begin{figure}[H]
    \centering
    \includegraphics[scale=1]{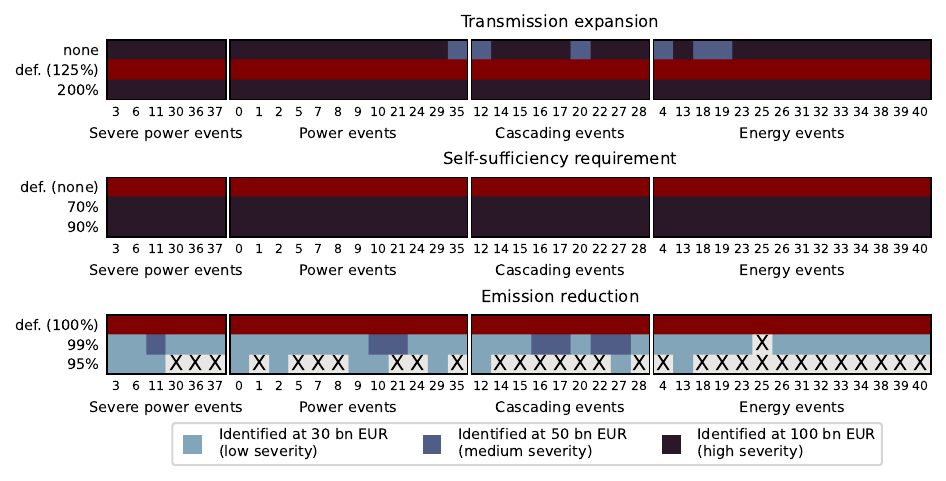}
    \caption{\textbf{Allowing marginal CO$_2$ emissions slashes severity of SDEs through usage of gas, but power deficits remain a concern}. Identification of SDEs across different system configuration varying assumptions on transmission expansion, equity requirements, CO$_2$ emission reductions. The colours indicate at which threshold (shades of blue; thresholds on accumulated costs based on shadow prices) the periods identified in the default scenarios (in red) are also identified in the sensitivity scenarios. Crosses (only for the emission reduction scenarios) indicate that these periods do not inflict significant cost on the system given the assumptions of the sensitivity scenario and hence are not identified as SDEs.}
    \label{fig:sensitivities}
\end{figure}

With more transmission in the European system, onshore wind becomes more profitable (versus a more local mix of solar, batteries, and distributed onshore wind; \cref{fig:affected-areas_no-trans}), increasing the exposure to the main driver of energy deficits. 
Still, increased transmission capacity balances wind generation daily by regional integration reducing the need for local balancing through fuel cells which are only needed during SDEs.
For extended transmission capacities, the same power deficit events are identified (\cref{fig:sensitivities}), only some energy deficit events at the margin change depending on the system configuration due to the altered generation patterns.

A shift away from offshore wind towards cheaper onshore wind and the usage of existing 195~GW of gas plants (which under the CO$_2$ constraint, can provide up to 2.5\% of annual generation) increases net load and dispatch of the resilience back-up significantly compared to fuel cells in the default scenario (\cref{fig:duration_fc_gas}).
Energy deficit events are no longer a major concern with the existing gas stock, but power deficit events are still identified at lower cost thresholds (\cref{fig:sensitivities}).

\begin{figure}[H]
    \centering
    \includegraphics[scale=1]{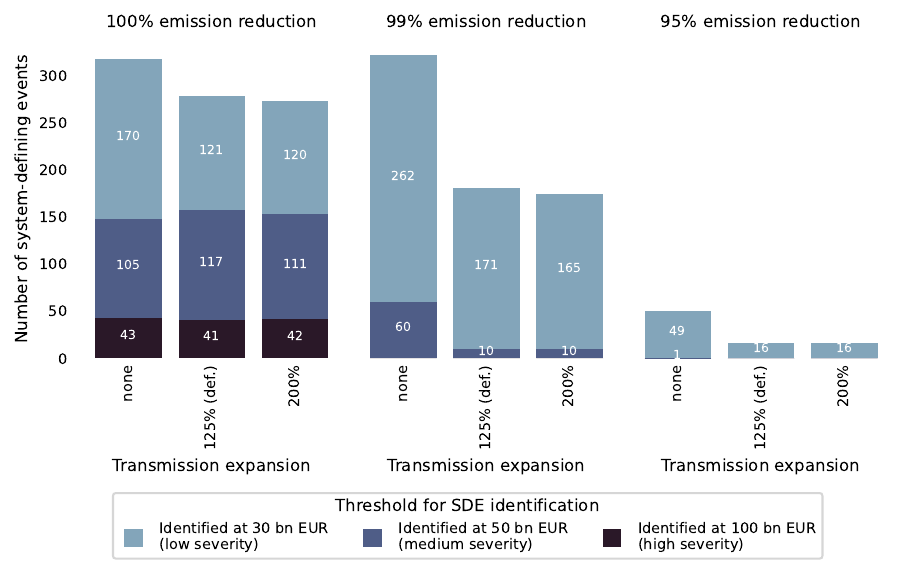}
    \caption{\textbf{Allowing marginal carbon emissions is more effective for SDEs than transmission expansion}. Number of identified system-defining events for different system configurations and thresholds. The varying thresholds indicate events of lower severity, i.e. 100 bn EUR is used to identify the 41 SDEs in the main paper, but a lower threshold of 50 bn EUR would have found 159=41+117 system-defining events. We do not explicitly show the scenarios with national self-sufficiency requirements, as the self-sufficiency constraints at 70\% and 90\% do not lead to any additional system-defining events.}
    \label{fig:sensitivity_number}
\end{figure}

\begin{figure}[H]
    \centering
    \includegraphics[scale=1]{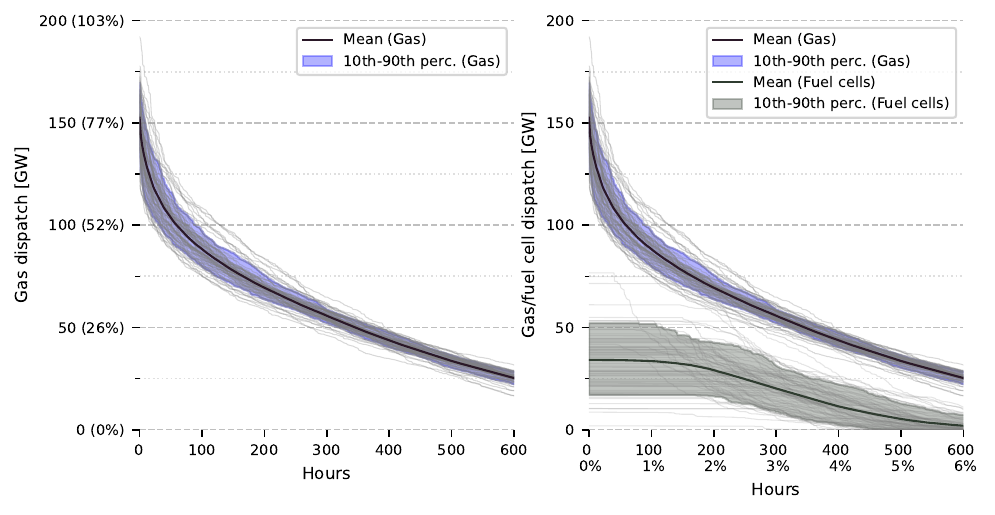}
    \caption{\textbf{Resilience back-ups (whether gas or fuel cells) have low capacity factors and use their full capacity very rarely (gas power plants barely at all), imperilling their viability}. Duration curves (one for each of the 80 weather years) for fuel cell discharge (as in Figure S4) in default scenario and gas power plants in sensitivity scenario. The range between the 10th and 90th percentile is shaded.}
    \label{fig:duration_fc_gas}
\end{figure}

\begin{figure}[H]
    \centering
    \includegraphics[scale=1]{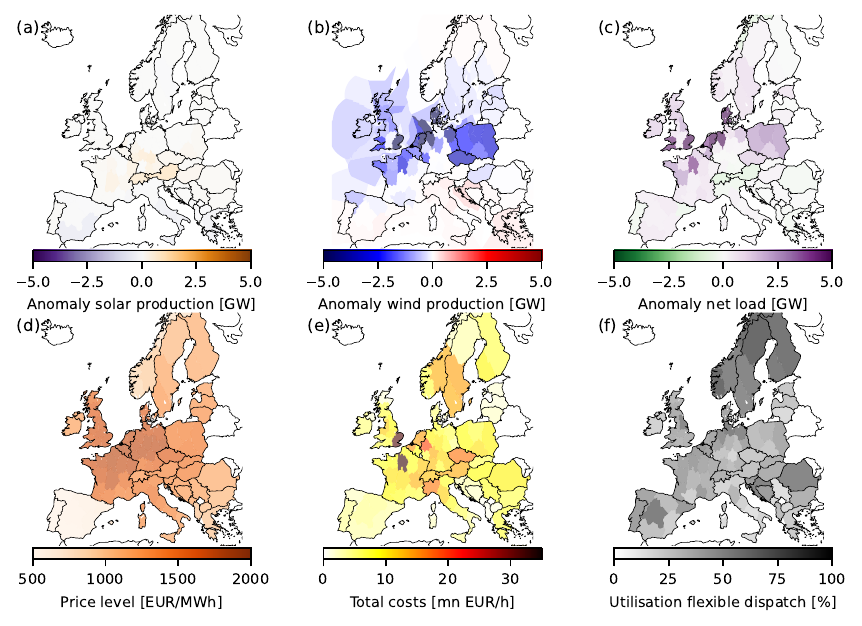}
    \caption{\textbf{With the current transmission network system stress spreads across all of Europe save for its peripheries}. Figure 4 but in the scenario without transmission expansion. Composites of power system impacts during the identified SDEs. The maps show the averages across all SDEs for anomalies of (a) solar generation, (b) wind generation, (c) net load, as well as (d) the average shadow price, (e) average hourly system costs, and (f) the utilisation of flexibility capacities defined as in Figure 3(a). Due to less transmission expansion to Scandinavia and Iberia, stress in the form of high shadow prices is contained more than with 25\% transmission expansion.}
    \label{fig:affected-areas_no-trans}
\end{figure}

Although transmission expansion significantly reduces system costs (already at the assumed 25\% expansion), the impact on system-defining events is less pronounced: more transmission does not reduce the frequency and only barely the severity of SDEs.
Transmission expansion is most beneficial for the system when excess wind power can be exported to the rest of the continent.
But that is exactly not the case during SDEs, which are often triggered by wind lulls around the North Sea where wind capacities are concentrated (Figure 4 (b) -- (c) and \cref{fig:affected-areas_no-trans} (b) -- (c)). 
Less impactful is the imposition of national self-sufficiency requirements: even when 90\% of annual generation needs to be sourced nationally, we identify the same SDEs (\cref{fig:sensitivities} and \cref{fig:sensitivity_number}).

\subsection*{Four types of system-defining events}

\begin{figure}[H]
    \centering
    \includegraphics[width=\linewidth]{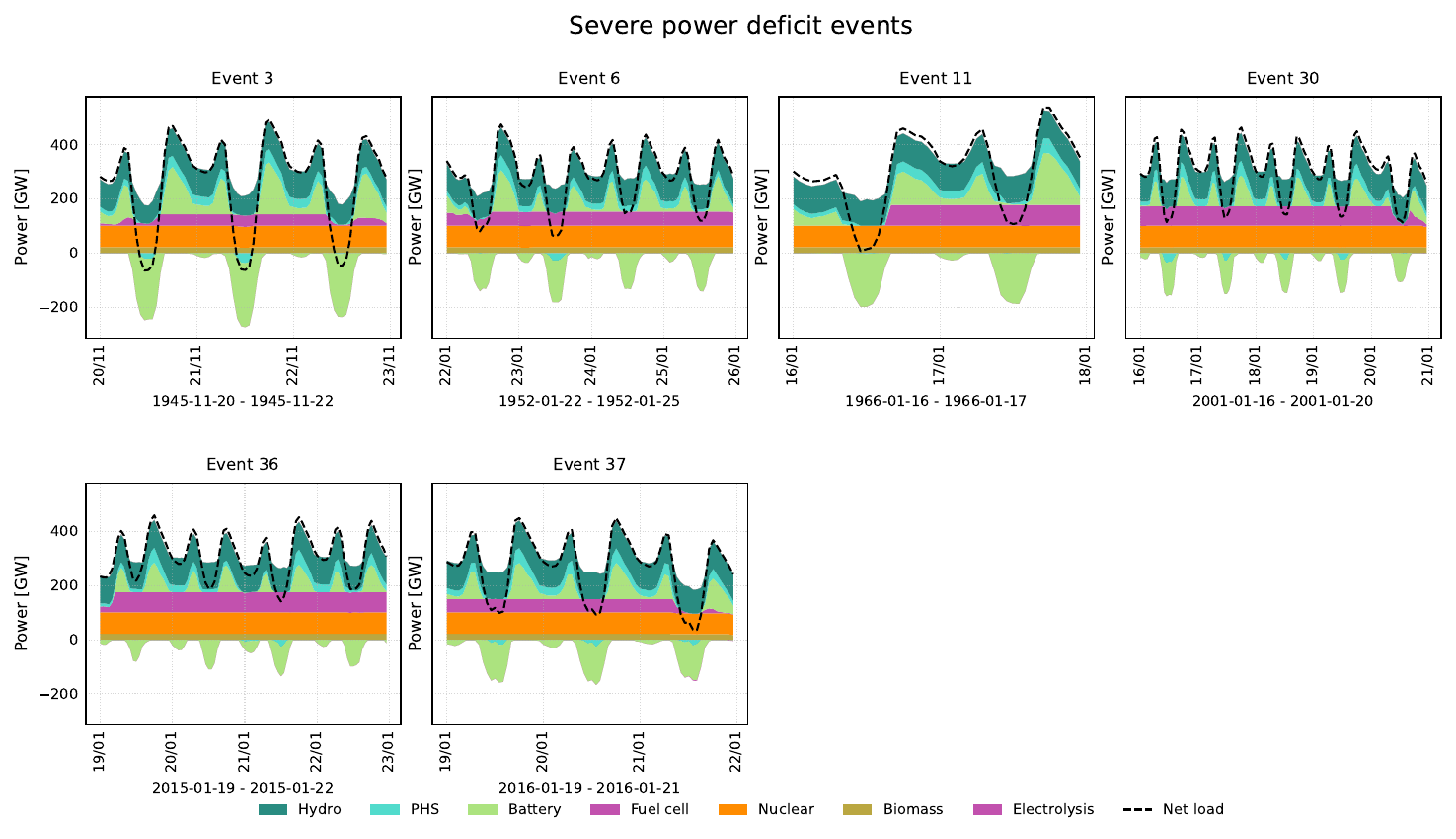}
    \caption{\textbf{Severe power deficit events are short and require large resilience back-ups}. Overview over cluster with severe power deficit events. Similar to Figure 2.}
    \label{fig:cluster_severe}
\end{figure}

\begin{figure}[H]
    \centering
    \includegraphics[width=\linewidth]{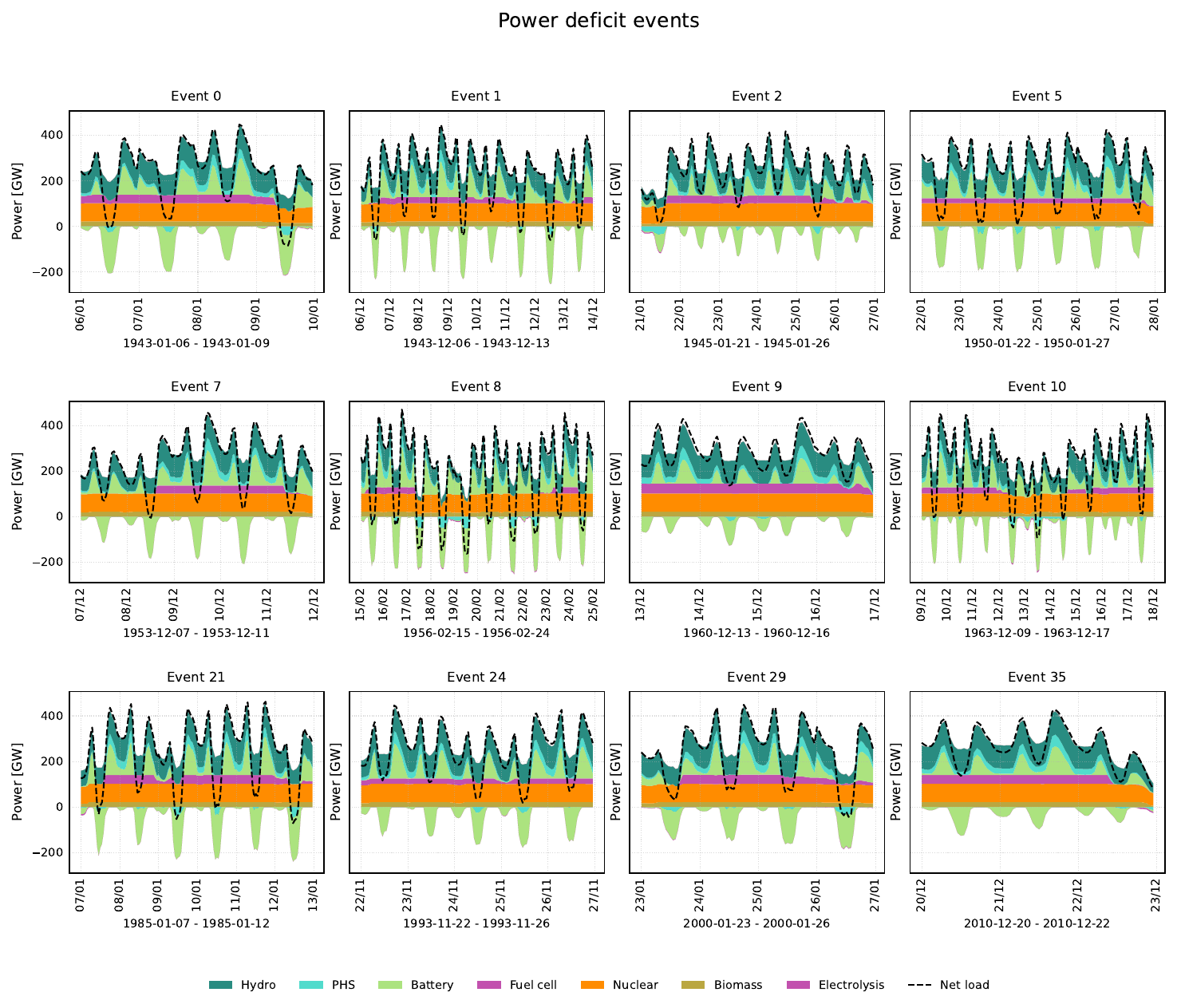}
    \caption{\textbf{Power deficit events are fairly short (but varying) and require resilience back-up to cover net load peaks}. Overview over cluster with  power deficit events. Similar to Figure 2.}
    \label{fig:cluster_power}
\end{figure}

\begin{figure}[H]
    \centering
    \includegraphics[width=\linewidth,]{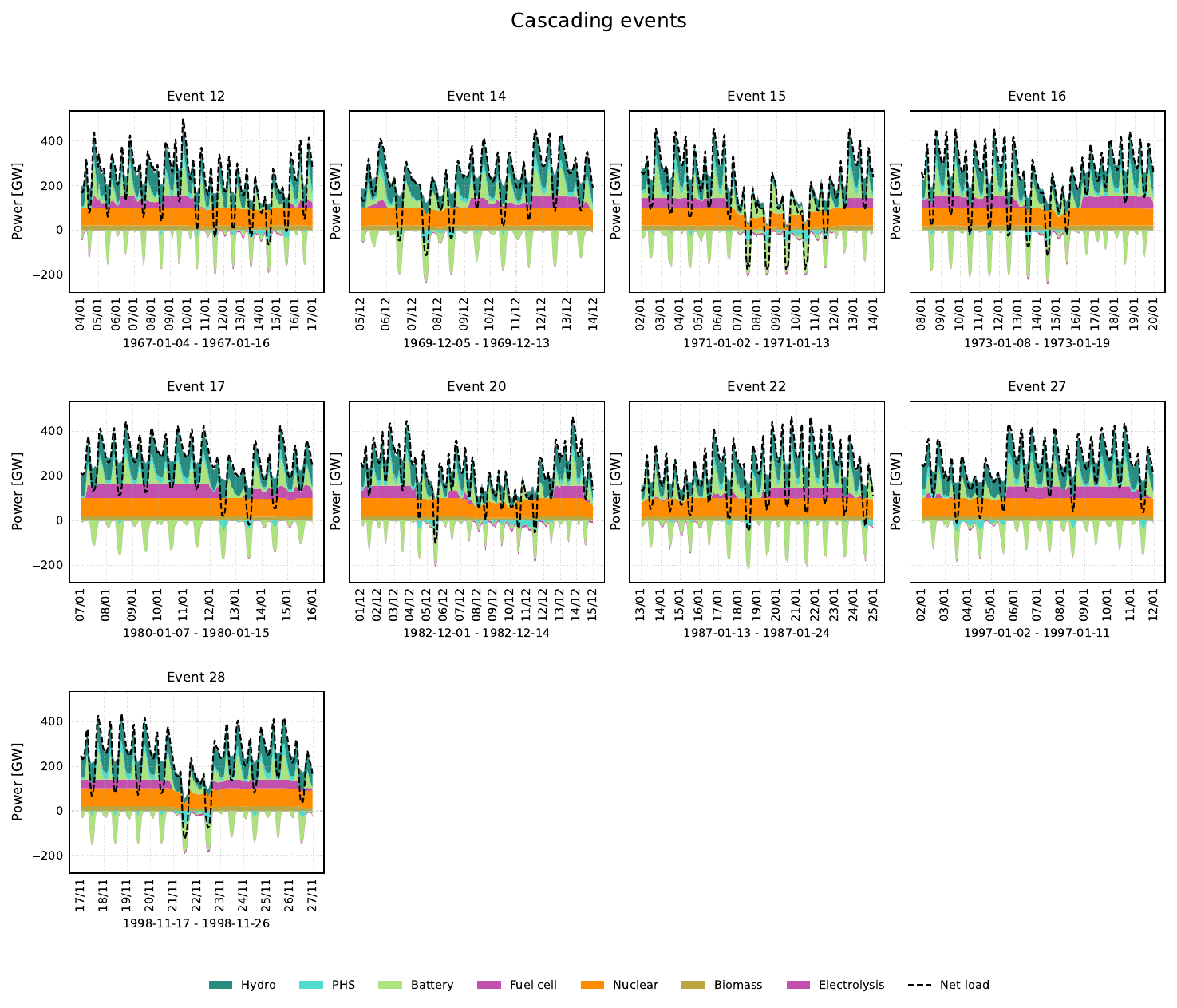}
    \caption{\textbf{Cascading events often consist of two power deficit periods with a short relief in between}. Overview over cluster with cascading events. Similar to Figure 2.}
    \label{fig:cluster_cascading}
\end{figure}

\begin{figure}[H]
    \centering
    \includegraphics[width=\linewidth,]{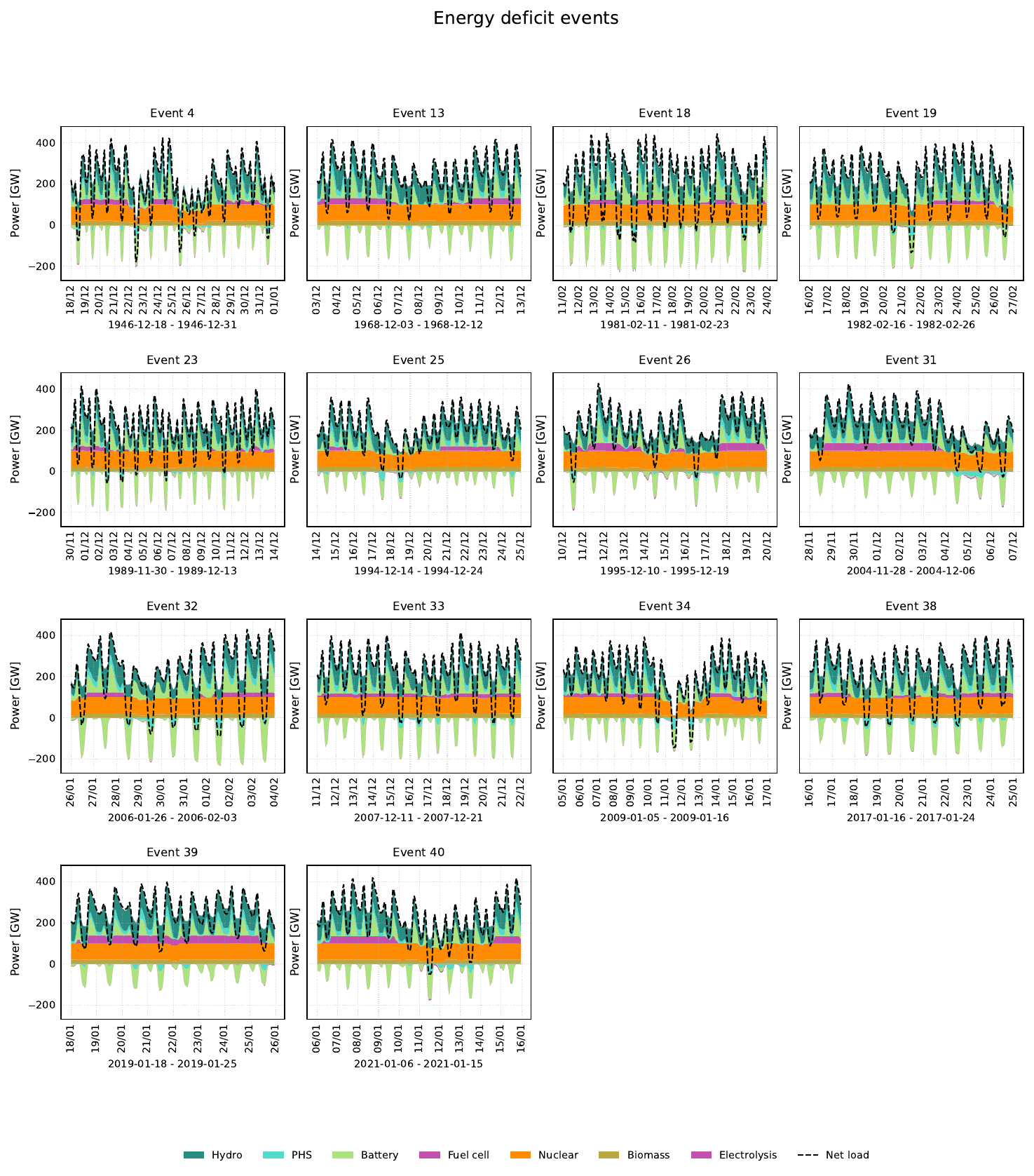}
    \caption{\textbf{Energy deficit events last the longest and need little resilience back-ups}. Overview over cluster with energy deficit events. Similar to Figure 2.}
    \label{fig:cluster_energy}
\end{figure}

\subsection*{Performance in short- and long-term resilience metrics}

\begin{figure}[H]
    \centering
    \includegraphics[scale=1]{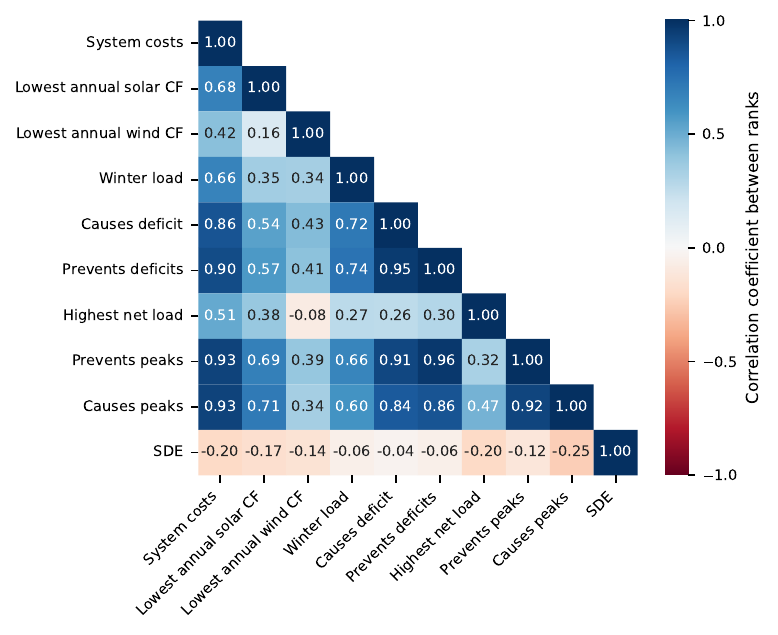}
    \caption{\textbf{Short-term resilience metrics such as occurrence of SDEs or net load do not necessarily correlate well with long-term resilience metrics like system costs}. Correlations for different resilience metrics.}
    \label{fig:corr_resilience_metrics}
\end{figure}

\begin{figure}[H]
    \centering
    \includegraphics[scale=1]{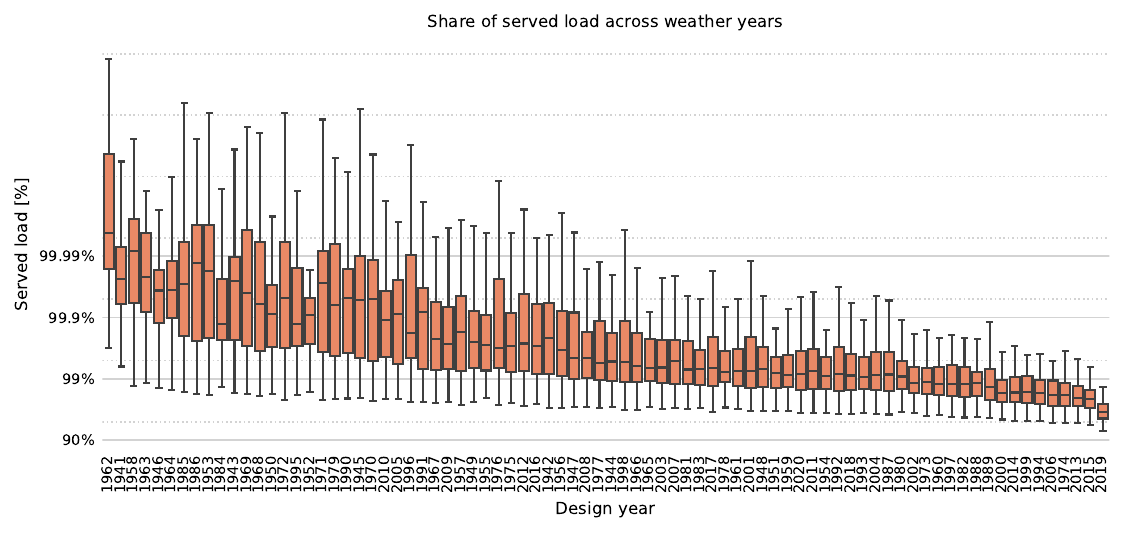}
    \caption{\textbf{System adequacy varies significantly across weather years}. Reliability of system designs across weather years. Interquartile ranges are marked and outliers are removed. Note the log-scale on the y-axis.}
    \label{fig:years_reliability}
\end{figure}


\begin{figure}[H]
    \centering
    \includegraphics[scale=1]{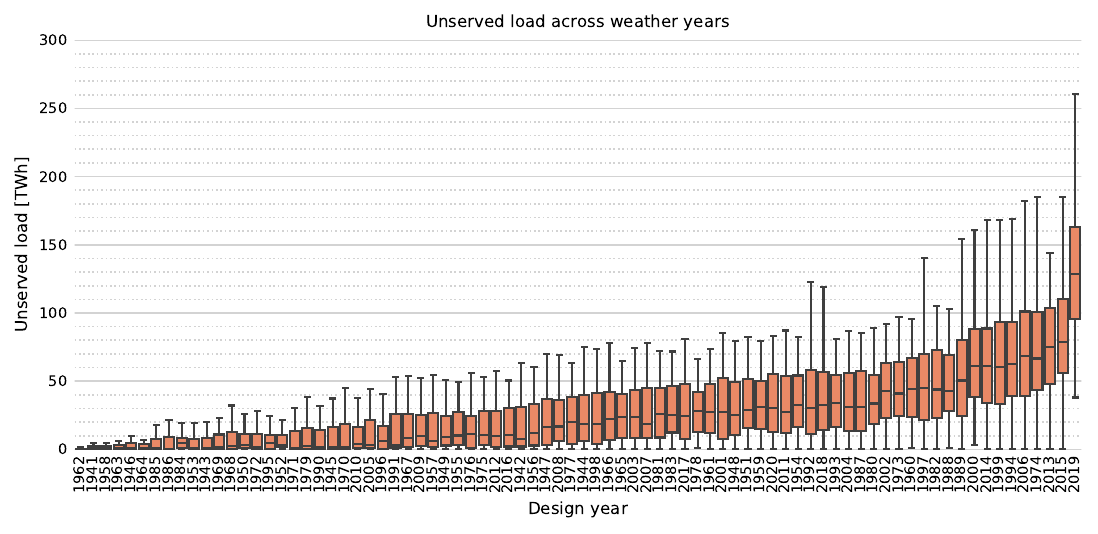}
    \caption{\textbf{Systems based on weather years like 2019/20 will be inadequate for unforeseen weather}. Unserved load for system designs across weather years. Interquartile ranges are marked and outliers are removed. Weather years with low unserved load are considered resilient to unforeseen weather.}
    \label{fig:years_unserved_design}
\end{figure}

\begin{figure}[H]
    \centering
    \includegraphics[scale=1]{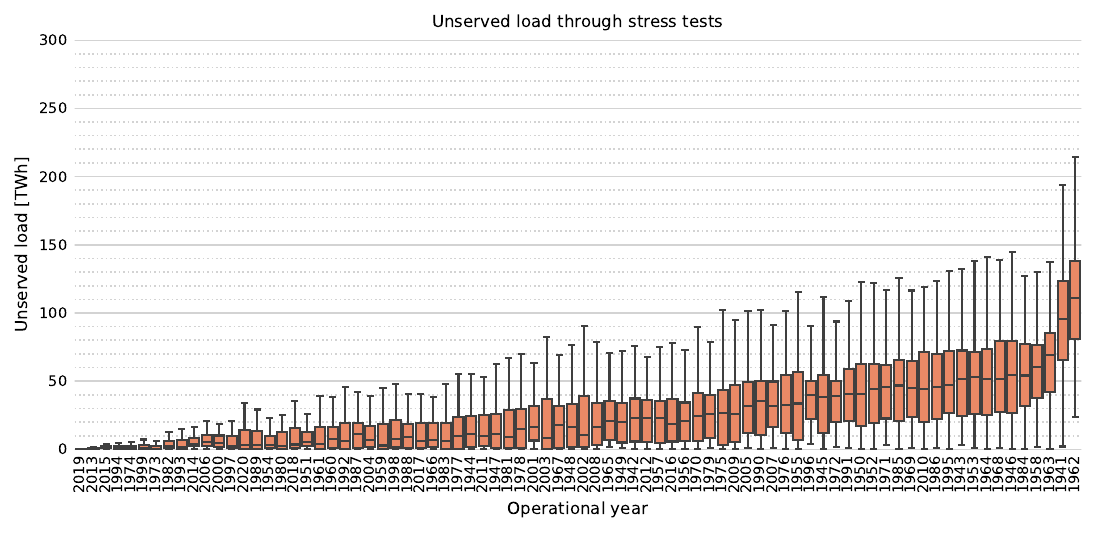}
    \caption{\textbf{Weather years such as 1941/2 or 1962/3 should serve as tests for long-term resilience}. Unserved load obtained during validation (dispatch optimisation) with a given weather year. Interquartile ranges are marked and outliers are removed. Weather years with high unserved load are considered hard to operate and thus suitable for stress tests.}
    \label{fig:years_unserved_operational}
\end{figure}

\begin{figure}[H]
    \centering
    \includegraphics[scale=1]{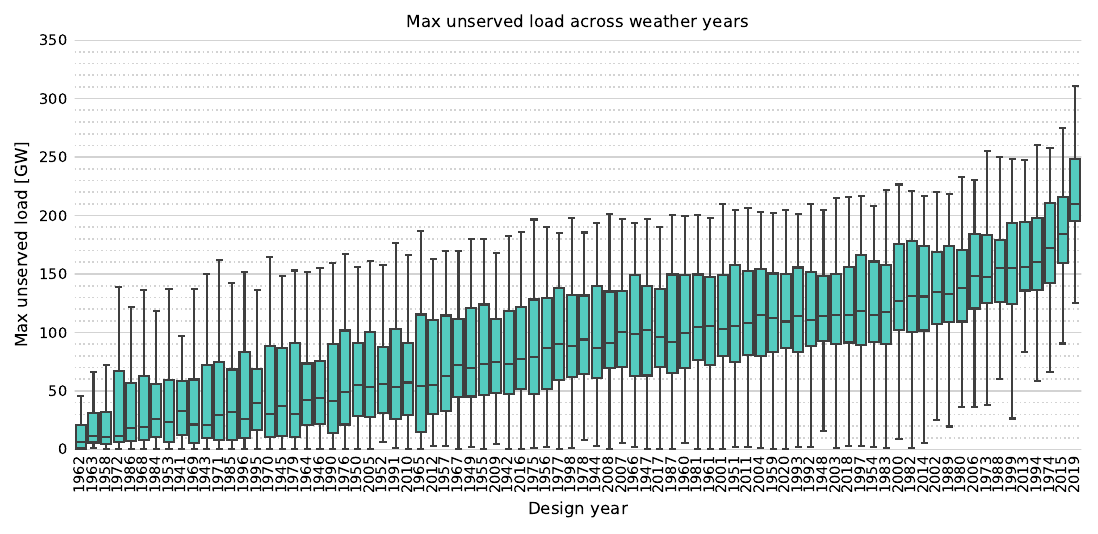}
    \caption{\textbf{Only a few design years have sufficient power capacities (e.g. resilience back-ups) to handle unforeseen weather reliably}. Maximal hourly unserved load for system designs across weather years. Interquartile ranges are marked and outliers are removed. Weather years with low maximal unserved load are considered resilient to short-term challenges.}
    \label{fig:years_shedding_design}
\end{figure}

\begin{figure}[H]
    \centering
    \includegraphics[scale=1]{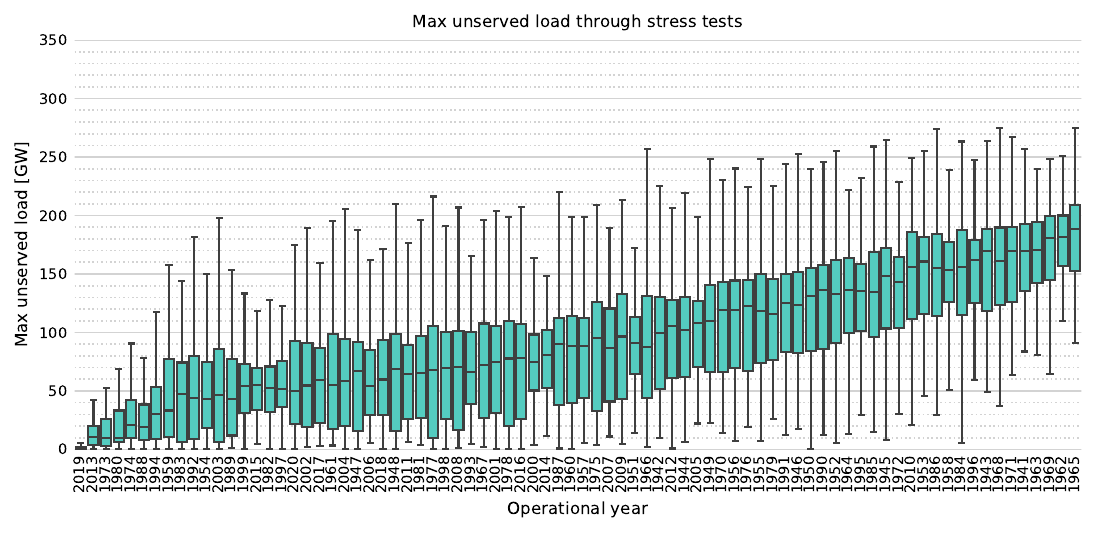}
    \caption{\textbf{Extreme cold (e.g. 1962/3) and wind droughts (e.g. 1965/6) can both serve to stress test resilience back-up capacities}. Maximal hourly unserved load obtained during validation (dispatch optimisation) with a given weather year. Interquartile ranges are marked and outliers are removed. Weather years with high maximal unserved load are considered hard to operate and thus suitable for stress tests on short-term resilience.}
    \label{fig:years_shedding_operational}
\end{figure}

\subsection*{Annual values of weather(-dependent) inputs (capacity factors and load)}

\begin{figure}[H]
    \centering
    \includegraphics[scale=1]{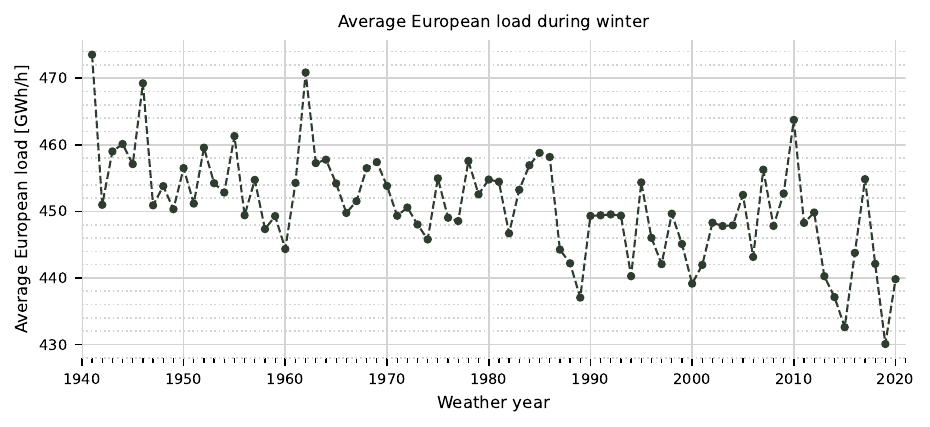}
    \caption{\textbf{Climate change signals are visible through decreasing load in the winter}. Average winter load across weather years.}
    \label{fig:winter_load}
\end{figure}

\begin{figure}[H]
    \centering
    \includegraphics[scale=1]{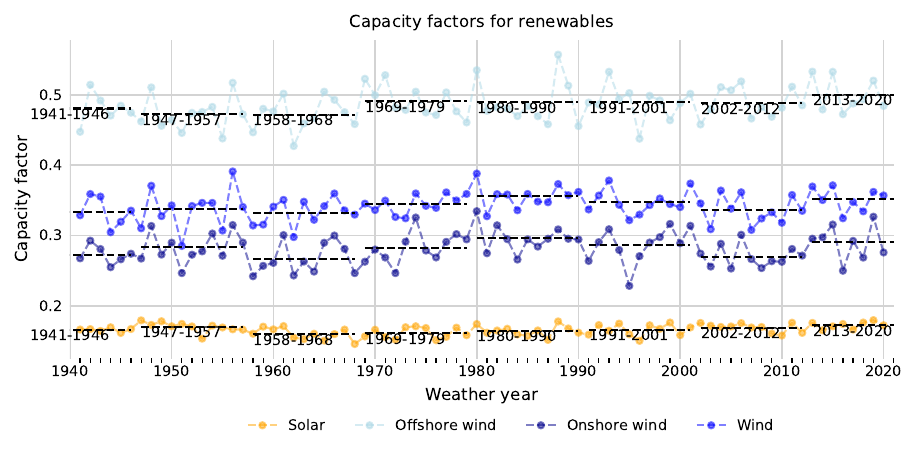}
    \caption{\textbf{Annual wind capacity factors vary across weather years with some decades having worse wind resources}. Annual capacity factors for different renewable resources (with 11-year averages marked to highlight the low wind resources between 1958 and 1968).}
    \label{fig:annual_decadal_cf}
\end{figure}


\subsection*{Similarity of different weather years}

\begin{figure}[H]
    \centering
    \includegraphics[scale=1]{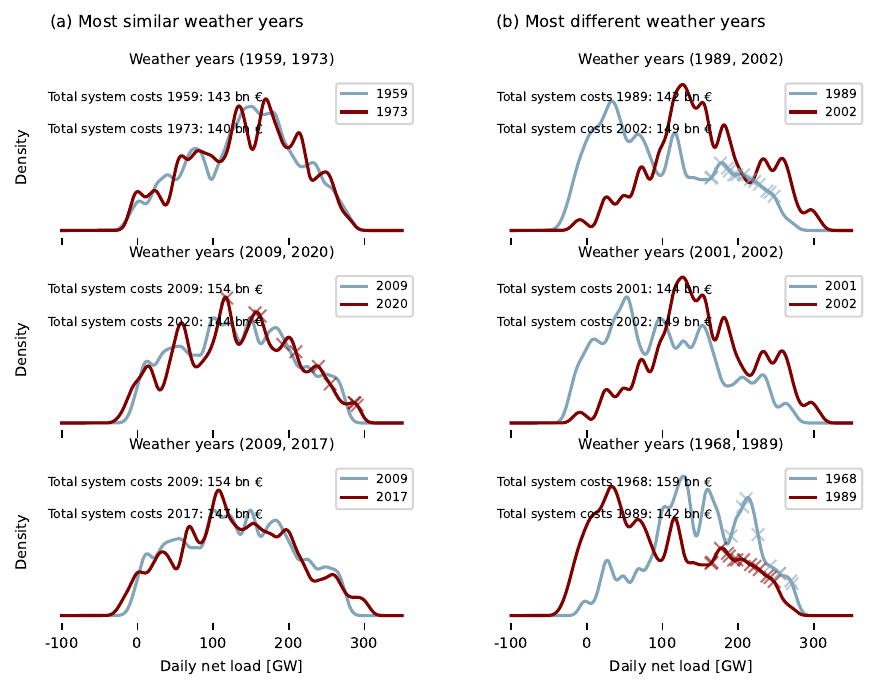}
    \caption{\textbf{Net load patterns can vary a lot between different weather years}. Similarity (based on Wasserstein metrics) of selected weather years based on daily net load with SDEs marked. Total system costs of the weather years is given as additional information.}
    \label{fig:wasserstein_net-load}
\end{figure}
\begin{figure}[H]
    \centering
    \includegraphics[scale=1]{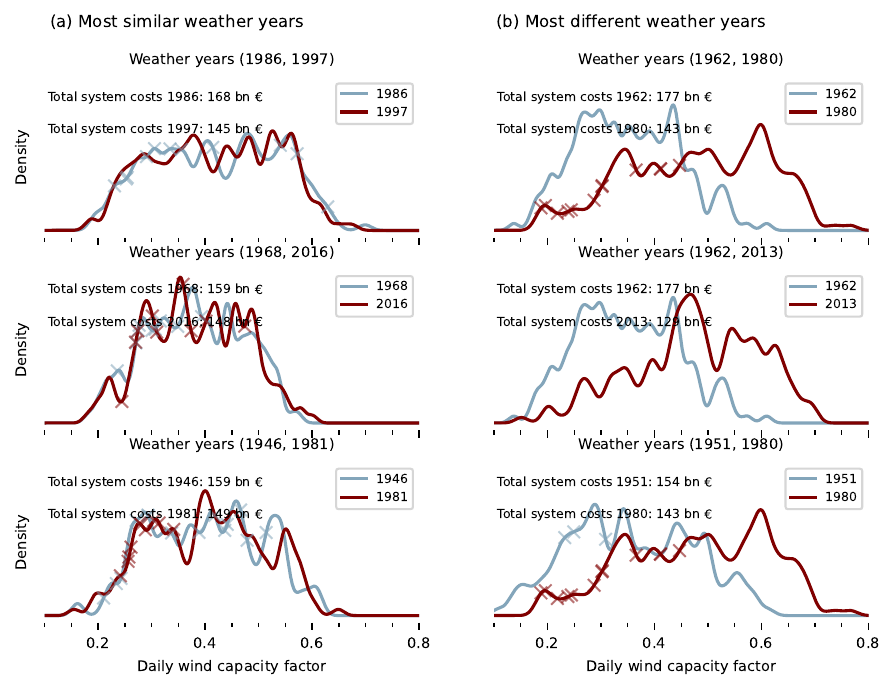}
    \caption{\textbf{Even very different years (in terms of wind) such as 51/52 and 80/81 can have ``similar'' SDEs}. Similarity (based on Wasserstein metrics) of selected weather years based on daily wind capacity factors with SDEs marked. Total system costs of the weather years is given as additional information.}
    \label{fig:wasserstein_wind}
\end{figure}